\documentclass[12pt,english]{article}
\usepackage[T1]{fontenc}
\usepackage[latin9]{inputenc}
\usepackage{float}
\usepackage{amsbsy}
\usepackage{amstext}
\usepackage{graphicx}
\usepackage{amssymb}
\usepackage{esint}

\makeatletter

\usepackage{epsfig}\@ifundefined{definecolor}
 {\usepackage{color}}{}
\usepackage{bm}\usepackage{latexsym}

\makeatother

\usepackage{babel}

\begin{document}


\title{Cold atoms in the presence of disorder}
\author{Boris Shapiro\\
Department of Physics, Technion-Israel Institute of Technology,\\
Haifa 32000, Israel}


\maketitle

\begin{abstract}
The review deals with the physics of cold atomic gases in the
presence of disorder. The emphasis is on the theoretical
developments, although several experiments are also briefly
discussed. The review is intended to be pedagogical, explaining
the basics and, for some of the topics, presenting rather detailed
calculations .

\end{abstract}




\section{Introduction}

The behavior of cold atoms and Bose-Einstein condensates (BEC) in
the presence of a random potential is an active area of research,
and several reviews on the subject had already appeared \cite{AI,SP1,Fall,Mod1,Lew,Cord}.
Those reviews discuss, among other things, the experimental work on
BEC spreading in a disordered wave guide, which have culminated in
the observation of one-dimensional Anderson localization \cite{Billy,Roati}.
The present review is devoted to two and three dimensional systems
of cold atoms, and it concentrates on topics not extensively covered
in the previous reviews. There exists considerable amount of theoretical
work on the subject, and recently few experimental results on spreading
of an atomic cloud, in two \cite{MR} and three \cite{Kondov,Jen}
dimensions, have been published. The review is intended to be self
contained and pedagogical, accessible to people without a solid background
in the physics of disorder. It therefore contains a significant amount
of background material.

Disorder is, of course, a very old theme in condensed matter physics,
and a lot of knowledge was accumulated over the years. Some of this
existing knowledge can be directly transferred to the field of cold
atoms, although there are also considerable differences between the
two fields. For one thing, a random potential for atoms is usually
created optically (an optical speckle) and 
it differs, in some respects, from the common models used in condensed
matter theory. Furthermore, in disordered electronic systems one is
usually interested in the nature of eigenstates at the Fermi level,
because this determines whether a given system is a metal or an insulator.
In contrast, a typical transport experiment with an atomic BEC involves
an expansion of the atomic cloud ( a \textquotedbl{}matter wave packet\textquotedbl{})
in the presence of disorder. Such a packet contains a broad spectrum
of components, some of which can propagate whereas the others get
localized. Nonlinear effects also come into play, especially at an
early stage of the expansion.

The equilibrium properties of a disordered BEC are also of interest,
in particular the phenomenon of the insulator-superfluid transition.
When the density of particles is low, they can at most form a fragmented
condensate, i.e. small disconnected lakes, with no coherent coupling
between them (an insulator). With an increase of the particle density
these lakes overlap and, at some critical density, a single coherent
condensate is formed (a superfluid).

It would hardly be possible to cover, in a single review, all aspects
of the broad and rapidly progressing field of disordered cold atom
systems. The choice of topics reflects, to some extent, personal preferences
of the author, and every effort was made to cover those topics with
some depth, clarity and precision. The emphasis is made on identifying
the relevant parameters and on discussing limiting cases, when simple
and clear physics emerges. Such simplified discussion might not be
sufficient for quantitative interpretation of the existing experiments
but, at least, it provides a good starting point for appreciating
the richness and complexity of the underlying physics.

Sections 2 and 3 of this review contain background material on speckle
potentials and on behavior of a quantum particle in a random potential.
Section 4 deals with a disordered BEC in equilibrium and discusses
the conditions for a disorder induced insulator-superfluid transition.
Section 5 is devoted to the dynamical problem of cold atoms expanding
in the presence of disorder. Both BEC and Fermi gas are treated. A
dynamical problem of a different type is considered in Section 6.
There a BEC with an initial phase modulation undergoes free expansion.
In the course of the expansion large density fluctuations can develop,
resulting in \textquotedbl{}matter wave caustics\textquotedbl{}. Some
concluding remarks are presented in Section 7.



\section{Optical speckles: A random potential for atoms}

An external potential (not necessarily random) for atoms can be created
by lasers \cite{String,Pet}. The laser radiation polarizes an atom
and changes its energy, e.g. the ground state energy, by an amount
\begin{eqnarray}
V({\bf r})=\frac{1}{2}\alpha(\omega)I({\bf r}),\label{s0}\end{eqnarray}
 where $I({\bf r})$ is the radiation intensity at the position ${\bf r}$
of the atom and $\alpha(\omega)$ is the atomic polarizability. The
latter depends on the radiation frequency $\omega$ and on the properties
of the atom, in particular, on the detuning $\delta=\omega-\omega_{R}$,
where $\omega_{R}$ is a resonance frequency of the relevant atomic
transition. The case $\delta>0$ (blue detuning) corresponds to a
positive (with respect to the unperturbed atomic ground state) potential,
while $\delta<0$ (red detuning) corresponds to a negative potential.
Thus, a random pattern of light intensity, referred to as speckle
pattern, will serve as a random potential for atoms. Such speckle
potentials are widely used in experiments with cold atoms \cite{Billy,MR,Kondov,Jen,Clement1,Clement2,Fort,Chen}.
Although there exists a rather comprehensive book on optical speckles
\cite{Goodman}, it is worthwhile to review the subject here, paying
attention also to the most recent work \cite{giglio,cerbino,gatti,magatti}.

\subsection{Qualitative discussion}

A schematic set up for making a speckle pattern is depicted in Fig.1.
\begin{figure}[H]
\centering{}\includegraphics[bb=0bp 0bp 792bp
612bp,scale=0.5]{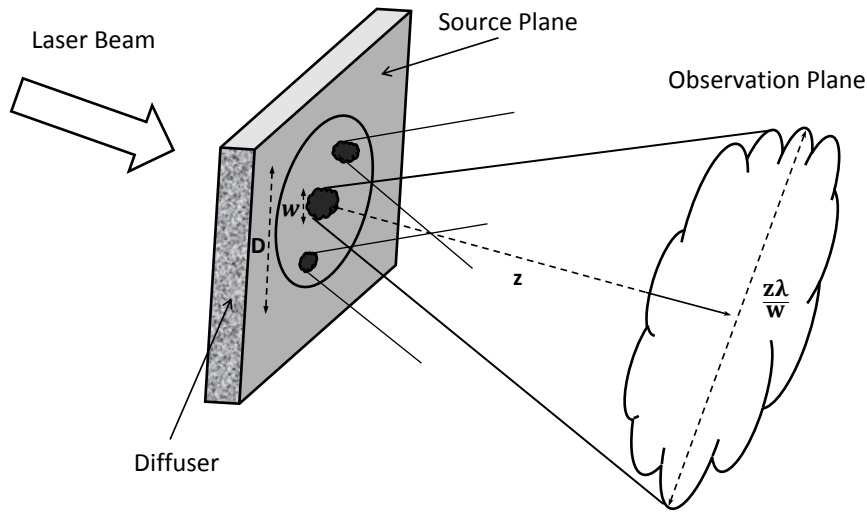}\caption{A laser beam, passing through a
diffuser, illuminates an area of size $D$ at the outgoing plane
$z=0$ (the source plane). Light from an elementary source of size
$w$ (three such sources are designated by black spots) propagates
in a cone of angular width $\lambda/w$ and illuminates a region of
size $z\lambda/w$ in the observation plane. Waves emerging from
different elementary sources interfere with each other, creating a
speckle pattern.}
\end{figure}
 A laser beam is transmitted through a diffuser. A common example
of a diffuser is a glass plate with a rough surface (ground
glass). Such a diffuser imprints a random phase on the transmitted
radiation, while the intensity at the outgoing plane (the plane
$z=0$) simply follows the shape of the beam, e.g., a Gaussian
laser beam. In the literature \cite{Goodman} one often studies a
geometry when a plane wave (the central part of a laser beam)
impinges on a circular or rectangular hole in a opaque screen, the
hole being covered by a ground glass diffuser. In this case the
light intensity is uniform within the corresponding circular or
rectangular domain, and it is zero outside that domain. The
detailed geometry of the set up is not essential for the
qualitative arguments.

Thus, an illuminated area, of size $D$, appears at the outgoing plane
of the diffuser (the $z=0$ plane, often referred to as source plane).
This illuminated area can be viewed as a collection of sources, of
a typical size $w$ (usually, of the order of the radiation wavelength
$\lambda$), with no spatial coherence between different sources.
The sources are monochromatic, of frequency $\omega$, and the $n'th$
source has a phase factor $e^{-i(\phi_{n}+\omega t)}$. Here $n$
runs from 1 up to the total number of sources, $N_{w}\sim(D/w)^{2}$,
and the phases $\phi_{n}$ change randomly from one source to another.
The collection of such random sources at the $z=0$ plane will produce
a random, three-dimensional intensity pattern (\textquotedbl{}grains\textquotedbl{}
of light) in the $z>0$ half-space. By choosing some distant observation
plane, $z=const$, one can study the transverse properties of such
speckle patterns. Their longitudinal properties, such as the typical
grain size in the $z$-direction, are also of interest. Although the
detailed theory of speckles is rather intricate, a rough qualitative
picture can be obtained by using a few basic facts about interference
and diffraction.

We discuss the transverse properties first. An elementary source,
of size $w$, in the source plane can be thought of as a particle
of this size which scatters the incoming light or, alternatively,
as a pinhole which diffracts the light. In either case the light will
propagate in a cone of angular width $\sim\lambda/w$, illuminating
a region of size $\sim z\lambda/w$ in the $z$-plane. Let's consider
first the case $(z\lambda/w)>D$, i.e. $z>(Dw/\lambda)\equiv z^{*}$.
This case corresponds to a \textquotedbl{}fully mixed\textquotedbl{}
speckle, when all elementary sources contribute to the intensity at
a typical point $P$ in the $z$-plane. 
The electric field at some point $P$ in the observation plane is
a sum of many complex amplitudes, with random phases, originating
from different areas at the source plane. As a result, the light intensity
$I$ across the observation plane is a random function of the position
$P$, with the characteristic bright and dark spots. An estimate of
the typical size of such speckle spot is analogous to the estimate
of the separation between neighboring maxima and minima in the standard
diffraction theory (the difference is that in the diffraction pattern
from, say, a circular hole maxima and minima occur only in the radial
direction, whereas in the speckle pattern they will be encountered
in all directions). Let us denote by $\alpha_{P}$ the phase difference
between the two waves arriving to the point $P$ from the opposite
edges of the illuminated area at the source plane. If the point $P$
is moved by a distance $s_{\bot}$, the phase difference $\alpha_{P}$
will change by $\sim s_{\bot}D/\lambda z$, i.e., a shift of the observation
point by a distance $s_{\bot}\sim\lambda z/D$ will change constructive
interference to destructive. The distance $s_{\bot}$ provides an
estimate for the size of one speckle spot. 
Note that our discussion pertains to what is called \textquotedbl{}a
free space geometry\textquotedbl{} \cite{Goodman}, when the light
propagates freely between the source plane and the observation plane.
In practice one usually employs the \textquotedbl{}imaging geometry\textquotedbl{}
\cite{Goodman}, when a converging lens is placed between the two
planes and the observation plane coincides with the focal plane of
the lens. The speckle spot size then becomes $s_{\bot}\sim\lambda f/D$,
where $f$ is the focal length of the lens, and the minimal size is
determined by the diffraction limit.

The picture outlined in the previous paragraph undergoes a
profound change when the distance $z$ to the observation plane
becomes smaller than $z^{*}$ - the so-called deep Fresnel zone
\cite{giglio,cerbino,gatti,magatti}.
According to the above estimate, for $z\sim z^{*}$ a speckle spot
reaches a size $s_{\bot}\sim w$. 
When $z$ decreases further, a point $P$ in the $z$-plane can receive
radiation only from a limited number of the elementary sources, namely,
the size of the region in the source plane which contributes to illumination
of point $P$ is $D_{eff}\sim\lambda z/w$. Using the above estimate
for the speckle spot size, with $D_{eff}$ instead of $D$, one obtains
$s_{\bot}\sim\lambda z/D_{eff}\sim w$. Thus, $s_{\bot}\sim w$ holds
starting from $z\sim z^{*}$ and all the way down to the minimal distance
$z\sim w^{2}/\lambda$.

We now briefly discuss the structure of a speckle pattern in the longitudinal
direction. Also in this direction the light intensity exhibits an
irregular sequence of maxima and minima, and separation between those
defines the typical size, $s_{\parallel}$, of the light \textquotedbl{}grain\textquotedbl{}
in the $z$-direction. Standard diffraction theory \cite{lipson}
tells us that a significant change of intensity in the $z$-direction
will occur when the corresponding phase, $D^{2}/\lambda z$, will
change by amount of order unity, i.e. such a change will occur over
a distance $s_{\parallel}\sim\lambda(z/D)^{2}$. In the deep Fresnel
zone, $z<z^{*}$, one has to substitute $D_{eff}$ for $D$, obtaining
\cite{gatti,magatti} $s_{\parallel}\sim w^{2}/\lambda$. 

\subsection{Analytic treatment}

Neglecting possible polarization effects, we consider a scalar
field which can be thought of as one particular component of a
monochromatic electric field. We denote the field at the source
plane by $\psi(\boldsymbol{\rho},0)$, where the vector
${\boldsymbol{\rho}}$ designates the in-plane position and the
second argument stands for $z=0$. The function
$\psi({\boldsymbol{\rho}},0)$ is significant only within the
region $|{\boldsymbol{\rho}}|\sim D$. The position in the
observation plane $z$ is denoted by ${\bf R}$. The field in this
plane created by the source field $\psi({\boldsymbol{\rho}},0)$,
is given by \cite{goodman1}:
\begin{eqnarray} \psi({\bf
R},z)=\int\frac{d^{2}q}{(2\pi)^{2}}\tilde{\psi}({\bf
q},0)\exp\left(i{\bf
q}\cdot{\boldsymbol{R}}+i\sqrt{k_{0}^{2}-q^{2}}z\right),\label{s1}\end{eqnarray}
 where $k_{0}=\omega/c=2\pi/\lambda$, and \begin{eqnarray}
\tilde{\psi}({\bf q},0)=\int d{\boldsymbol{\rho}}\psi({\boldsymbol{\rho}},0)e^{-i{\bf q}\cdot\boldsymbol{\rho}}.\label{s2}\end{eqnarray}
 The expression (\ref{s1}) for $\psi({\bf R},z)$ is exact, containing
both propagating and evanescent waves, and it has a simple
interpretation. For $z=0$ the source is Fourier expanded with
respect to the transverse coordinate $\boldsymbol{\rho}$, and the
factor $\exp(i\sqrt{k_{0}^{2}-q^{2}}z)$ is the
\textquotedbl{}Fourier propagator\textquotedbl{} which relates the
transverse Fourier transform at the $z$-plane, $\tilde{\psi}({\bf
q,z)}$, to $\tilde{\psi}({\bf q,0)}$.

In the paraxial approximation the Fourier propagator is
approximated by $\exp[ik_{0}z-(iq^{2}z/2k_{0})]$. Substituting
(\ref{s2}) into (\ref{s1}), with the approximated Fourier
propagator, and performing integration over ${\bf q}$ leads to:
\begin{eqnarray} \psi({\bf R},z)=\frac{k_{0}}{2\pi
iz}e^{ik_{0}z}\int
d\boldsymbol{\rho}\psi(\boldsymbol{\rho},0)\exp\left[i\frac{k_{0}}{2z}({\bf
R-\boldsymbol{\rho})^{2}}\right],\label{s3}\end{eqnarray}
 which is the mapping of the source field, $\psi(\boldsymbol{\rho},0)$,
onto the field $\psi({\bf R},z)$ at the observation plane.

$\psi(\boldsymbol{\rho},0)$ is a random function of the position
$\boldsymbol{\rho}$. It is described by the correlation function
$\overline{\psi(\boldsymbol{\rho},0)\psi^{*}(\boldsymbol{\rho}',0)}$,
where the overbar indicates a statistical average. Eq. (\ref{s3})
enables one to immediately write down the correlation function for
the bulk speckle pattern,
$\overline{\psi(\boldsymbol{R},z)\psi^{*}(\boldsymbol{R}',z')}$.
We give the explicit expression only for the
\textquotedbl{}transverse\textquotedbl{} correlation function
($z=z'$): \begin{eqnarray}
\overline{\psi(\boldsymbol{R},z)\psi^{*}(\boldsymbol{R}',z)}=
\left(\frac{k_{0}}{2\pi z}\right)^{2}\int
d\boldsymbol{\rho}d\boldsymbol
{\rho}'\overline{\psi(\boldsymbol{\rho},0)\psi^{*}(\boldsymbol{\rho}',0)}\exp\left[i\frac{k_{0}}{z}(\triangle\boldsymbol{R}-\triangle\boldsymbol{\rho})\cdot(\boldsymbol{R}_{0}-\boldsymbol{\rho}_{0})\right],\label{s4}\end{eqnarray}
 where $\triangle\boldsymbol{R}\equiv\boldsymbol{R}-\boldsymbol{R}'$,
$\triangle\boldsymbol{\rho}\equiv\boldsymbol{\rho}-\boldsymbol{\rho}'$,
and $\boldsymbol{R}_{0}\equiv\frac{1}{2}(\boldsymbol{R}+\boldsymbol{R}')$,
$\boldsymbol{\rho}_{0}\equiv\frac{1}{2}(\boldsymbol{\rho}+\boldsymbol{\rho}'$).

This expression simplifies if one assumes a white noise source,
i.e.
$\overline{\psi(\boldsymbol{\rho},0)\psi^{*}(\boldsymbol{\rho}',0)}=\chi\overline{I}(\boldsymbol{\rho})\delta(\boldsymbol{\rho}-\boldsymbol{\rho}')$,
where $\chi$ is a proportionality constant and
$\overline{I}(\boldsymbol{\rho})=\overline{|\psi(\boldsymbol{\rho},0)|^{2}}$
is the average intensity at point $\boldsymbol{\rho}$ in the
source plane. For this case one obtains \cite{Goodman}:
\begin{eqnarray}
\overline{\psi(\boldsymbol{R},z)\psi^{*}(\boldsymbol{R}',z)}=\chi\left(\frac{k_{0}}{2\pi
z}\right)^{2}
\exp\left(i\frac{k_{0}}{z}\boldsymbol{R}_{0}\cdot\triangle\boldsymbol{R}\right)\int
d\boldsymbol{\rho}\overline{I}(\boldsymbol{\rho})\exp\left(-i\frac{k_{0}}{z}\triangle\boldsymbol{R}\cdot\boldsymbol{\rho}\right),\label{s5}\end{eqnarray}
 which is proportional to the Fourier transform of $\overline{I}(\boldsymbol{\rho})$.
If $\overline{I}(\boldsymbol{\rho})$ significantly differs from
zero in a domain of size $D$, then the correlation function
$\overline{\psi(\boldsymbol{R},z)\psi^{*}(\boldsymbol{R}',z)}$
decays on a characteristic scale $s_{\bot}\sim\lambda z/D$, which
defines the typical transverse size of a speckle spot, as
estimated in Sec. 2.1. For simple cases, when
$\overline{I}(\boldsymbol{\rho})$ is constant within an aperture
of a rectangular or circular shape, the integral in (\ref{s5}) can
be evaluated explicitly \cite{Goodman}. Another simple and
instructive example is
$\overline{I}(\boldsymbol{\rho})=I_{0}\exp(-\rho^{2}/2D^{2})$. The
integral in (\ref{s5}) is then given by $2\pi
D^{2}\exp(-\frac{D^{2}k_{0}^{2}\triangle R^{2}}{2z^{2}})$ which,
again, reveals the characteristic decay length $\lambda z/D$.

One can relax the assumption of a white noise source and consider
a more general correlation function, at the source plane
\cite{gatti}: \begin{eqnarray}
\overline{\psi(\boldsymbol{\rho},0)\psi^{*}(\boldsymbol{\rho}',0)}=C(\triangle\boldsymbol{\rho})\overline{I}(\boldsymbol{\rho}_{0}),\label{s6}\end{eqnarray}
 where the function $C$ decays on a shot scale, $w$, whereas the
average intensity $\overline{I}$ changes on a much larger scale,
$D$. Using this expression as an input in Eq. (\ref{s4}), one can
derive the transverse properties of the speckle pattern. The term
$k_{0}(\boldsymbol{\rho}\cdot\triangle\boldsymbol{\rho})/z\sim k_{0}Dw/z$,
in the exponent of Eq. (\ref{s4}), can be neglected for $z>Dw/\lambda\equiv z^{*}$.
In this case the double integral in (\ref{s4}) factorizes into a
product of two integrals which are the Fourier transforms of $I(\boldsymbol{\rho}_{0})$
and $C(\triangle\boldsymbol{\rho})$. These Fourier transforms determine,
respectively, the speckle spot size $s_{\bot}\sim\lambda z/D$ and
the extent $\lambda z/w$ of the entire speckle pattern in the $z$-plane.
In the opposite case, $z<z^{*}$, the speckle pattern \textquotedbl{}freezes\textquotedbl{},
as described in Section 2.1. 

The more general correlation function,
$\overline{\psi(\boldsymbol{R},z)\psi^{*}(\boldsymbol{R}',z')}$,
with $z\neq z'$, has been studied in detail in
\cite{gatti,magatti}, where the normalized function,
$\gamma(\boldsymbol{R},z;\boldsymbol{R}',z')\equiv\overline{\psi(\boldsymbol{R},z)\psi^{*}(\boldsymbol{R}',z')}/\sqrt{\overline{I}(\boldsymbol{R},z)\overline{I}(\boldsymbol{R}',z')}$
was introduced (this function is denoted there by $\mu$). Simple
analytic expressions for
$|\gamma(\boldsymbol{R},z;\boldsymbol{R}+\triangle\boldsymbol{R},z+\triangle
z)|^{2}$ can be obtained if one chooses in (\ref{s6})
$C(\triangle\boldsymbol{\rho})=\exp[-(\triangle\boldsymbol{\rho}/w)^{2}]$
and
$\overline{I}(\boldsymbol{\rho}_{0})=I_{0}\exp[-(\boldsymbol{\rho}_{0}/D)^{2}]$
\cite{gatti,magatti}. In certain limiting cases the expression
\begin{eqnarray}
\left|\gamma(\boldsymbol{R},z;\boldsymbol{R}+\triangle\boldsymbol{R},z+\triangle
z)\right|^{2}=\frac{1}{1+(\triangle
z/s_{\parallel})^{2}}\exp\left[-\frac{2(\triangle\boldsymbol{R}/s_{\perp})^{2}}{1+(\triangle
z/s_{\parallel})^{2}}\right]\label{corr gamma}\end{eqnarray}
 emerges. In the deep Fresnel zone $s_{\parallel}=\pi w^{2}/\lambda$
and $s_{\perp}=w$ (Eq. (12) in \cite{magatti}), whereas in the
Fresnel zone (and for $\triangle z$ not too large)
$s_{\parallel}=\lambda z^{2}/\pi D^{2}$ and $s_{\perp}=\lambda
z/\pi D$ \cite{Note1}. The correlation function (\ref{corr gamma})
is quite unusual: not only is it  strongly anisotropic but also
the effective correlation length in the transverse direction,
$s_{\perp eff}\sim s_{\perp}\sqrt{1+(\triangle
z/s_{\parallel})^{2}}$ , depends on the separation $\triangle z$
between the points in the longitudinal direction. Moreover,
$s_{\perp eff}$ increases with $\triangle z$, so that long range
correlations in the transverse direction develop.

A simpler three-dimensional potential, corresponding to the
interference pattern in an ergodic cavity, was considered in
\cite{Kuhn}. It is fully isotropic and the square of the
correlation function is \begin{eqnarray}
\gamma^{2}(\boldsymbol{r})=sinc^{2}(2\pi r/\lambda),\label{corr
gamma1}\end{eqnarray}
 where $\lambda$ is the light wave length. 

Let us recall that the potential acting on the atoms is proportional
not to the field but to its intensity, see Eq.(\ref{s0}). Therefore,
denoting $(\boldsymbol{R},z)\equiv\boldsymbol{r}$, the correlation
function for the potential, $\overline{V(\boldsymbol{r})V(\boldsymbol{r}')}$,
is proportional to $\overline{I(\boldsymbol{r})I(\boldsymbol{r}')}=\overline{\psi(\boldsymbol{r})\psi^{*}(\boldsymbol{r})\psi(\boldsymbol{r}')\psi^{*}(\boldsymbol{r}')}$,
which requires averaging of a product of four fields. 
For a Gaussian random function the latter reduces to the sum of
two possible contractions of the fields, with the result
\cite{Goodman}:
\begin{eqnarray}
\overline{I(\boldsymbol{r})I(\boldsymbol{r}')}=\overline{I}(\boldsymbol{r})\overline{I}(\boldsymbol{r}')+|\overline{\psi(\boldsymbol{r})\psi^{*}(\boldsymbol{r}')}|^{2}.\label{s7}\end{eqnarray}
Since $\overline{I}(\boldsymbol{r})$ changes on a scale much larger
than the decay length of the correlation function, one can set $\overline{I}(\boldsymbol{r})=\overline{I}(\boldsymbol{r}')\equiv I_{0}$.
Rewriting (\ref{s7}) in terms of the potential then yields \cite{Kuhn,Mini}:
\begin{eqnarray}
\overline{V(\boldsymbol{r})V(\boldsymbol{r}')}=V_{0}^{2}[1+|\gamma(\boldsymbol{r}-\boldsymbol{r}')|^2],\label{s8}\end{eqnarray}
 where $V_{0}$ is the average value of the potential at point $\boldsymbol{r}$
and $\gamma=\overline{\psi(\boldsymbol{r})\psi^{*}(\boldsymbol{r}')}/I_{0}$
is the normalized correlation function of the speckle field. It follows
from (\ref{s8}) that the second moment $\overline{V^{2}}=2V_{0}^{2}$.
The $n'th$ moment, $\overline{V^{n}}$, will involve averaging of
$n$ pairs of fields and thus $n!$ contractions, giving $\overline{V^{n}}=n!V_{0}^{n}$.
This implies that the probability distribution of the potential value,
at some point $\boldsymbol{r}$, is given by the Rayleigh distribution
\begin{eqnarray}
P(V)=\frac{1}{V_{0}}\exp(-V/V_{0}),\label{s9}\end{eqnarray}
 where a blue detuned potential has been assumed (for red detuning
both $V$ and $V_{0}$ are negative, so that (\ref{s9}) should be
written with a minus sign). The Rayleigh distribution for the
intensity is a direct consequence of the fact that the speckle
field - being a sum of many contributions with random phases -
obeys the central limit theorem, i.e., the real and the imaginary
part of the field are Gaussian distributed \cite{Goodman}. The
distribution $P(V)$ is not symmetric with respect to its average
value $V_{0}$, and it is convenient to introduce the deviation
$\delta V=V-V_{0}$, with the correlation function \begin{eqnarray}
\overline{\delta V(\boldsymbol{r})\delta
V(\boldsymbol{r}')}=V_{0}^{2}\Gamma\left(\frac{\boldsymbol{r}-\boldsymbol{r}'}{R_{0}}\right),\label{s10}\end{eqnarray}
 where $|\gamma|^{2}$ was denoted by $\Gamma$ and the argument $(\boldsymbol{r}-\boldsymbol{r}')$
was rescaled by the correlation decay length $R_{0}$. Computer generated
images of a speckle potential (in $2d$) can be found, e. g., in \cite{Kuhn,Pezze}.

Finally, let us mention that making an optical speckle  is not the
only way to create a random potential for atoms. Another option is
to introduce into the system some foreign,
\textquotedbl{}impurity\textquotedbl{} atoms, trapped in a random
fashion at the nodes of an optical lattice \cite{Gavish}. These
impurity atoms form a frozen, random pattern which scatters the
mobile atoms of the system under investigation. This method for
creating a random potential has been implemented in the recent
study of a one-dimensional Bose system \cite{Gad}.

\section{Quantum particle in a random potential}

The behavior of a quantum particle in a random potential has been
extensively studied in condensed matter physics \cite{ES,L,Lee,Kram,Akker,Efetov,Mirlin}.
The theoretical tools range from diagrammatic technique \cite{Akker},
to self-consistent approach \cite{Voll}, to field theory methods
\cite{Efetov,Mirlin}. The most popular model is that of a Gaussian
random potential, with or without correlations (the latter case is
known as the \textquotedbl{}white-noise limit\textquotedbl{}). This
model features not only in condensed matter physics but also in the
field of cold atoms, for instance, in the study of the disorder induced
superfluid-insulator transition \cite{Shk,Fal}. The Gaussian potential
differs in some respects from a speckle potential. The latter is bounded
from below (blue detuning) or from above (red detuning). This will
clearly lead to large changes in the particle behavior close to the
boundary of the spectrum. In addition, the speckle potentials lack
symmetry (in statistical sense, of course), with respect to the average
value, so that products of an odd number of $\delta V$'s do not average
to zero. Therefore the diagram technique for a speckle potential \cite{Kuhn,Mini}
differs somewhat from the standard one \cite{Akker}. To a large extent,
however, there are no qualitative differences, as long as one is in
the weak scattering regime. In particular, the mean free path, calculated
in the Born approximation, is \textquotedbl{}universal\textquotedbl{},
in the sense that it only depends on the correlation function (\ref{s10})
but not on the detailed statistics of the potential.




Consider a random potential $V(\boldsymbol{r})$, with some average
value $\overline{V}$.
This
value is included into the unperturbed Hamiltonian, and the perturbation
is $\delta V(\boldsymbol{r})$. Thus, the bottom of the unperturbed
spectrum is $\overline{V}$ and, if the particle energy $\epsilon$
is reckoned from this point, the unperturbed spectrum is simply $\epsilon=\hbar^{2}k^{2}/2m$.
With this caveat, most of the ensuing discussion applies to a broad
class of potentials, including speckle and Gaussian potentials. 
Viewing the potential relief from the \textquotedbl{}$\overline{V}$-level\textquotedbl{},
one can envisage a collection of potentials barriers and wells, of
typical magnitude $V_{0}$ and typical size $R_{0}$. It is convenient
to define the \textquotedbl{}correlation energy\textquotedbl{} \begin{eqnarray}
E_{0}=\frac{\hbar^{2}}{2mR_{0}^{2}},\label{cor.energy}\end{eqnarray}
 where $m$ is the particle's mass. $E_{0}$ has the clear meaning
of a zero point energy for a particle confined to a spatial region
of size $R_{0}$. 
There are, thus, three relevant energy scales in our problem - the
typical variation $V_{0}$ of the random potential, the correlation
energy $E_{0}$ and the particle energy $\epsilon$ - and the
physics depends on the two ratios: $V_{0}/E_{0}$ and
$\epsilon/E_{0}=(kR_{0})^{2}$. In particular, $kR_{0}<<1$
corresponds to short range correlations, when the correlation
length $R_{0}$ is shorter than the de Broglie
wavelength 
of the particle. The opposite case, $kR_{0}>>1$, describes smooth
(on the wavelength scale) 
 disorder, for which a semiclassical approximation might be a good
starting point. The space dimension, $d$, is also an important factor.
For $d=1,2$ all eigenstates (in an infinite system) are localized,
while for $d=3$ there is a critical energy $E_{c}$ which separates
between extended and localized states. In this review we concentrate
on two and three dimensions.

The value of the parameter $(V_{0}/E_{0})\equiv\eta$ is crucial for
understanding the transport picture. For $\eta<<1$ a typical well
is too weak to bind a particle. In this regime semiclassical considerations
are not appropriate and classical percolation is of no relevance whatsoever
for understanding the quantum localization transition. In the opposite
case, $\eta>>1$, there is a broad region in the particle energy spectrum,
where semiclassical considerations do apply. The ultimate transition
to localization is, of course, of quantum nature also for this case
but now the transition occurs close (somewhat above) to the classical
percolation threshold and an interesting crossover between classical
percolation and Anderson localization takes place. All this is discussed
below in some detail. Before making various estimates and calculations,
it is essential to specify whether $\eta$ is large or small, and
it is convenient to separately discuss the following cases:

\subsection{$d=3$ and $\eta<<1$}

A particle of sufficiently high energy will be weakly scattered by
the random potential and will propagate by diffusion. 
When the energy of the particle is lowered, the scattering becomes
stronger and quantum interference effects cease to be negligible.
These effects lead to Anderson localization \cite{And} below some
energy $E_{c}$, called the mobility edge.

Before trying to estimate $E_{c}$ let us make the following
remark: The $3d$ speckle potential is not
\textquotedbl{}generic\textquotedbl{} in the sense that the
integral $\int\Gamma(\textbf{r})d^{3}r$ diverges, due to the
long-range character of the correlation function
$\Gamma(\textbf{r})$ (see equations (\ref{corr gamma}) and
(\ref{corr gamma1})). In contrast, the impurity potential,
mentioned at the end of Sec. 2.2, belongs to the standard type of
a \textquotedbl{}generic\textquotedbl{} potential, usually
encountered in condensed matter physics. The expression for the
mean free path, and therefore an estimate for $E_{c}$, depends on
whether the potential is \textquotedbl{}generic\textquotedbl{} or
not. In the forthcoming discussion the
\textquotedbl{}generic\textquotedbl{} case will be considered
first and, then, some necessary changes for
the speckle potential will be pointed out. 

Let us now estimate 
the characteristic energy near which the diffusive transport breaks
down. 
Qualitatively, the random potential can be viewed as a collection
of scatterers, of strength $V_{0}$ and size $R_{0}$. Since $V_{0}<<E_{0}$,
a single scatterer can be treated within the Born approximation, regardless
of the particle energy. One should, however, distinguish between \textquotedbl{}slow\textquotedbl{}
particles, with energies $\epsilon<<E_{0}$ (i.e. $kR_{0}<<1$) and
\textquotedbl{}fast\textquotedbl{} particles, satisfying $\epsilon>>E_{0}$
(i.e. $kR_{0}>>1$). For slow particles the scattering cross-section
is estimated as \cite{LL} $\sigma\sim\eta^{2}R_{0}^{2}$. 
Since the number of scatterers per unit volume is $n_{scat}\sim R_{0}^{-3}$,
the mean free path \begin{eqnarray}
l=\frac{1}{\sigma n_{scat}}\sim\frac{1}{\eta^{2}}R_{0}\sim\left(\frac{\hbar^{2}}{m}\right)^{2}\frac{1}{V_{0}^{2}R_{0}^{3}}.\label{mfp}\end{eqnarray}

When one refers to \textquotedbl{}a quantum particle\textquotedbl{},
one actually has in mind a wave packet with a well defined wave vector
$\textbf{k}$, such that the uncertainty $\triangle k\lesssim k$.
The minimal size of such a packet is of the order of $k^{-1}$ and,
for the above reasoning to be consistent, this must be much smaller
than $l$, which leads to the requirement $kl>>1$. The condition
$kl\sim1$ (the Ioffe-Regel criterion) corresponds to a characteristic
energy \begin{eqnarray}
W\sim V_{0}\eta^{3}\sim\left(\frac{m}{\hbar^{2}}\right)^{3}\left(V_{0}^{2}R_{0}^{3}\right)^{2}.\label{W}\end{eqnarray}
 Thus, when the particle energy $\epsilon$ (counted from the average
value of the potential) decreases and approaches $W$, the diffusive
transport breaks down and quantum interference effects come into play
in an essential way . The potential strength and range appear in Eqs.
(\ref{mfp},\ref{W}) only in the combination $V_{0}^{2}R_{0}^{3}$.

We now return to the fast particles, with $\epsilon>>E_{0}$. Such
high energy particles scatter primarily in the forward direction,
within a cone of an angular width $\triangle\theta<1/kR_{0}$.
Therefore the scattering cross section for fast particles is
$\sigma\sim\eta^{2}/k^{2}$ \cite{LL} (smaller by a factor
$(1/kR_{0})^{2}$ than the cross section for slow particles).
Furthermore, the cross section relevant for transport is not
$\sigma$ but the transport cross section,
$\sigma^{*}\sim\sigma/(kR_{0})^{2}$. The small extra factor,
$(1/kR_{0})^{2}$, accounts for the fact that in a single
scattering event the particle direction changes only by a small
angle, randomly distributed in a cone of angular width
 $\Delta\theta\sim1/kR_{0}$. Therefore it will take $(kR_{0})^{2}$ scattering events to
completely randomize the initial direction. Thus, the transport
mean free path is:
\begin{eqnarray} l^{*}=\frac{R_{0}^{3}}{\sigma^{*}}\sim
R_{0}\left(\frac{\epsilon}{V_{0}}\right)^{2}.\label{l'}\end{eqnarray}

The qualitative estimate in Eqs. (\ref{mfp}), (\ref{l'}) can be
supported by a calculation based on a diagrammatic perturbation theory
\cite{Akker}. In that theory the transport mean free time $\tau^{*}$
is related to the correlation function of the potential. In the leading
order in $V_{0}^{2}$, the inverse mean free time (the scattering
rate) is expressed in terms of the Fourier transform $\tilde{\Gamma}(\textbf{q})$
of the correlation function $\Gamma(\frac{\boldsymbol{R}}{R_{0}})$,
defined in Eq. (\ref{s10}): \begin{eqnarray}
\frac{1}{\tau^{*}}=\frac{2\pi}{\hbar}V_{0}^{2}\int\frac{d^{3}k'}{(2\pi)^{3}}\tilde{\Gamma}
(\textbf{k}-\textbf{k}')\delta\left(\epsilon-\frac{\hbar^{2}k'^{2}}{2m}\right)(1-\cos\theta)\nonumber \\
=\frac{\pi}{\hbar}V_{0}^{2}\nu\int_{0}^{\pi}d\theta
\sin\theta\tilde{\Gamma}\left(2k\sin\frac{\theta}{2}\right)(1-\cos\theta),\label{inverse}\end{eqnarray}
 where $\epsilon=\frac{\hbar^{2}k^{2}}{2m}$ is the particle energy
and $\nu=mk/2\pi^{2}\hbar^{2}$ is the density of states (per spin)
at that energy. The argument $2k\sin\frac{\theta}{2}$ corresponds
to the momentum transfer for (elastic) scattering at an angle
$\theta$ and the factor $(1-\cos\theta)$ accounts for the
difference between the scattering and the transport cross
sections. Taking, for example, a Gaussian correlation function,
$\Gamma=\exp(-R^{2}/R_{0}^{2}$), performing the integral and
writing the result in terms of $l^{*}=v\tau^{*}$ ( $v=\hbar k/m$
is the particle velocity), yields \begin{eqnarray}
l^{*}=\frac{2R_{0}}{\sqrt{\pi}}\left(\frac{\epsilon}{V_{0}}\right)^{2}\frac{1}{1-(1+k^{2}R_{0}^{2})\exp(-k^{2}R_{0}^{2})}.\label{l''}\end{eqnarray}
 For $kR_{0}<<1$ one can expand the exponent, obtaining $l^{*}=4R_{0}/\eta^{2}\sqrt{\pi}$
(compare to the qualitative estimate in (\ref{mfp})), whereas in
the opposite case the result is simply $l^{*}=\frac{2R_{0}}{\sqrt{\pi}}(\frac{\epsilon}{V_{0}})^{2}$
(compare with (\ref{l'})). Eq.(\ref{l''}) can be re-written in terms
of the diffusion coefficient $D_{\epsilon}=vl^{*}/3$. For the limiting
cases the result is: \begin{equation}
D_{\epsilon}=\left\{ \begin{array}{c}
\frac{\hbar}{m}\frac{4}{3\surd\pi}\left(\frac{\epsilon-\overline{V}}{W}\right)^{1/2}\;\;\;\;\;,\;\;\; W<<(\epsilon-\overline{V})<<E_{0}\\
(\frac{8V_{0}R_{0}^{2}}{\pi m})^{1/2}\left(\frac{\epsilon-\overline{V}}{V_{0}}\right)^{5/2}\;\;,\;\;\;\;\;\;\;\;\;\epsilon>>E_{0}\end{array}\right.\label{D}\end{equation}
 Note that we have restored $\overline{V}$ in these expressions,
to give them the more convenient form, \textquotedbl{}invariant\textquotedbl{}
with respect to the choice of the reference energy. Other types of
the correlation function $\Gamma(\textbf{R})$ can be considered in
a similar way. The order-of magnitude estimates (\ref{mfp}), (\ref{l'})
do not depend on the specific shape of the correlation function, provided
that $\int\Gamma(\textbf{R})d^{3}R$ is finite.

Thus, for the case $\eta<<1$ the Born approximation is valid in a
broad range of energies, the only condition being $(\epsilon-\overline{V})>>W$.
When $(\epsilon-\overline{V})$ approaches $W$, the Born approximation
breaks down and the diffusion coefficient becomes of the order of
$\hbar/m$. This is the natural unit of \textquotedbl{}quantum diffusion\textquotedbl{}
(analogous to the quantum conductance $e^{2}/\hbar$) and it signals
a crossover to a purely quantum mode of transport. At $E_{c}$ the
diffusion coefficient drops to zero. Somewhat above $E_{c}$ it obeys
a power law, $D_{\epsilon}\propto(\epsilon-E_{c})^{\nu}$, with $\nu\approx1.57$
\cite{Slevin}. Below $E_{c}$ all states are localized. The exact
position of $E_{c}$ depends on the type of the random potential.
It should be located somewhere in the vicinity of the average potential
value $\overline{V}$, although the precise location, and even the
sign (with respect to $\overline{V}$), is not known. The aforementioned
Ioffe-Regel criterion marks breakdown of the diffusive transport but
it does not enable one to determine the location of $E_{c}$. In technical
terms, the mean free path is related to the imaginary part of the
self-energy, whereas $E_{c}$ can be affected also by the real part,
which causes a shift of the spectrum. The real part, calculated in
the self-consistent Born approximation (see \cite{Skip}, \cite{Tig}
and Sec. 5.3), turns out to be negative, of the order of $\eta V_{0}$,
which indicates that $E_{c}$ is below $\overline{V}$. One should
keep in mind, however, that the self-consistent Born approximation
is not really a controlled one.

For energies well below $E_{c}$, when $(E_{c}-\epsilon)>>W$, one
enters the region of strongly localized states- the
\textquotedbl{}Lifshitz tail\textquotedbl{}. While for high
energies the density of states $\nu(\epsilon)$ is close to that of
a free particle, in the Lifshitz tail $\nu(\epsilon)$ is small and
it depends on the type of the potential. For a Gaussian potential
the tail was studied long ago, by the method of optimal
fluctuation \cite{H,Z,L}. At extremely low energies,
$(E_{c}-\epsilon)>E_0$, the optimal fluctuation is just a single
well, of size $R_{0}$ and depth $\sim|\epsilon|$, which
immediately leads to $\ln\nu(\epsilon)\propto-\epsilon^{2}/V_0^2$.
For energies $W<<E_c-\epsilon<<E_{0}$ the optimal fluctuation is
of different nature. In this energy interval a single well of size
$R_{0}$ and depth $\sim|\epsilon|$ cannot bind a particle, and the
optimal fluctuation corresponds to a much broader well, of radius
$R_{\epsilon}\sim(\hbar^{2}/m|\epsilon|)^{1/2}>>R_{0}$. Since such
a well is comprised of $\sim(R_{\epsilon}/R_{0})^{d}$
\textquotedbl{}elementary\textquotedbl{} wells, the corresponding
probability is $\exp(-\epsilon^{2}/V_{0}^{2}$), raised to power
$(R_{\epsilon}/R_{0})^{d}$. This yields

\begin{eqnarray}
\ln\nu(\epsilon)\propto-\left(\frac{\epsilon^{2}}{V_{0}^{2}}\right)\left(\frac{\hbar^{2}}{m|\epsilon|R_{0}^{2}}\right)^{d/2}\sim-\left(\frac{|\epsilon|}{E_{t}}\right)^{2-(d/2)}\;\;\;\;\;,\;\;\;
E_{t}=\left(\frac{V_{0}^{4}}{E_{0}^{d}}\right)^{1/(4-d)}.\label{dos}\end{eqnarray}
 In $3d$, $E_{t}=W$, and in $2d$ it is the characteristic crossover
energy $V_{0}^{2}/E_{0}$, which will be identified in Sec. 3.3. It
is interesting to note that the same energy scale $W$ (in $3d$)
emerges both from the Ioffe-Regel criterion and from the analysis
of the Lifshitz tail. The factor $\sim(R_{\epsilon}/R_{0})^{d}$,
entering the derivation of (\ref{dos}), can be also interpreted in
a different way: Since the spatial extent of the localized wave
function, $R_{\epsilon}$, strongly exceeds the typical variation
length $R_{0}$ of the random potential, the particle experiences
not the original potential but an effective Gaussian potential
with a variance smaller by a factor $(R_{\epsilon}/R_{0})^{d}$.

In $3d$, (\ref{dos}) can be written as
$\ln\nu(\epsilon)\sim-\mathcal{L}/R_{\epsilon}$, where
$\mathcal{L}\sim R_{0}/\eta^{2}$ is a characteristic length,
specifying the short range disorder, and termed the Larkin length
in \cite{Fal} (it is of the same order of magnitude as the mean
free path in the weak scattering regime, (\ref{mfp})). The
localized states in the Lifshitz tail are rare, separated by
exponentially large distances, and the tunneling amplitude between
a pair of such states is exponentially small. When $\epsilon$
approaches the mobility edge, by a few $E_{t}$, the size of the
localized wave function approaches $\mathcal{L}$, the overlap
becomes significant and the tunneling amplitude is of the order of
unity: the concept of the optimal fluctuation ceases
to be relevant, 

So far there have been no quantitative studies of the Lifshitz tails
for speckle potentials in $2d$ and $3d$ (the $1d$ case was considered
in \cite{fal1}). For a red-detuned potential, which is not bounded
from below, one can argue that 
 Eq. (\ref{dos}) for the Lifshitz tail is still valid. Although the
original potential is not Gaussian (see (\ref{s9})), the particle
localized in a optimal fluctuation feels an effective potential,
obtained by integrating the original,
\textquotedbl{}microscopic\textquotedbl{} potential over a volume
$R_{\epsilon}^{d}$. This results in an effective potential with an
approximately Gaussian statistics, with a variance by the factor
$(R_{\epsilon}/R_{0})^{d}$ smaller than that of the original
potential, so that (\ref{dos}) is recovered. This consideration
does not hold in the extreme tail, $|\epsilon|<E_{0}$, when the
optimal fluctuation is just a single well of depth
$\sim|\epsilon|$ and size $\sim R_{0}$. The density of states then
follows the tail of the potential distribution (\ref{s9}), i.e.
$\ln\nu(\epsilon)\propto-|\epsilon|/V_0$.

For the blue-detuned speckle potential the picture is different.
Here one can expect the validity of Eq. (\ref{dos}) only if the
particle energy satisfies $W<<(E_{c}-\epsilon)<<V_{0}$, i.e. when
$\epsilon$ is well below $E_{c}$ but still far away from the
bottom of the spectrum. When $\epsilon$ approaches the bottom of
the spectrum, $\epsilon=0$ (here it is natural to reckon the
energy from the bottom of the spectrum), a new tail, of different
nature emerges. Close to $\epsilon=0$ the density of states can be
estimated as follows \cite{L} (the argument applies in any
dimension $d$). In order to arrange for a state with a small
energy, one has to \textquotedbl{}clean out\textquotedbl{} a
region of size $R_{\epsilon}\sim(\hbar^{2}/m\epsilon)$, i.e. no
barriers of height larger than $\epsilon$ should be allowed in
that region. Such a region consists of many,
$\sim(R_{\epsilon}/R_{0})^{d}$, independent elements, of size
$R_{0}$ each. The probability that the potential in a given
element does not exceed $\epsilon$ is $\sim\epsilon/V_{0}$.
Therefore, for $\epsilon\rightarrow 0$
\begin{eqnarray}
\nu(\epsilon)\propto\left(\frac{\epsilon}{V_{0}}\right)^{(R_{\epsilon}/R_{0})^{d}}\;\;\Rrightarrow\;\;
\ln\nu(\epsilon)\propto
-\left(\frac{E_{0}}{\epsilon}\right)^{d/2}ln\frac{V_{0}}{\epsilon}
\label{dos1}\end{eqnarray}
The two expressions, (\ref{dos}) and
(\ref{dos1}) should match somewhere half way between the bottom of
the spectrum and $E_{c}$.


As has been stated in the beginning of this section, for the 3d
speckle potential the integral $\int\Gamma(\textbf{r})d^{3}r$
diverges, so that the expression (\ref{mfp}) for the mean free
path and the subsequent estimate for $W$ do not apply. The formula
(\ref{inverse}), however, does apply and, for the ergodic cavity
potential (\ref{corr gamma1}), it yields for the mean free path
(in the small $k$ limit, when there is no difference between $l$
and $l^*$) \cite{Kuhn}:
\begin{eqnarray}
l=\left(\frac{\hbar^{2}}{m}\right)^{2}\frac{k}{\pi V_{0}^{2}R_{0}^{2}},\label{mfp'}\end{eqnarray}
 with $R_{0}=\lambda/2\pi$. Thus, for $kR_{0}<<1$, the mean free
path is proportional to $k$, unlike the $k$-independent result in
(\ref{mfp}). 
The condition $kl\sim1$ will now give a characteristic energy
$V_{0}^{2}/E_{0}$, instead of (\ref{W}). This energy provides an
estimate for $E_c$, with respect to $\overline{V}$.

Finally, the study of transport in a $3d$ anisotropic speckle
potential, with correlation function (\ref{corr gamma}) (or,
possibly, even some more general and complicated cases
\cite{gatti,magatti}) only begins and the first work on the
subject has just appeared \cite{Pira}. Due to the anisotropy,
diffusion is described by a tensor which, in the weak scattering
regime, can be computed with the help of the Fourier transform of
(\ref{corr gamma}): \begin{eqnarray}
\tilde{\Gamma}(\textbf{q}_{\bot},q_{\|})=\pi\sqrt{2\pi}\;\frac{s_{\bot}s_{\|}}{q_{\bot}}\exp\left[-\frac{q_{\bot}^{2}s_{\bot}^{2}}{8}-2\left(\frac{q_{\|}s_{\|}}{q_{\bot}s_{\bot}}\right)^{2}\right]\;,\label{anisotropic}\end{eqnarray}
 where $\textbf{q}_{\bot}$ and $q_{\|}$ are the transverse and the
longitudinal (i.e. $z$) component of the transmitted momentum. Because
of the long range transverse correlations in the speckle potential
(see the discussion after Eq. (\ref{corr gamma})) $\tilde{\Gamma}(\textbf{q}_{\bot},q_{\|})$
has a peculiar essential singularity at $q_{\bot}=0$. More theoretical
work is needed in order to understand how such strongly anisotropic,
\textquotedbl{}non-generic\textquotedbl{} potentials affect the standard
picture of Anderson localization. It is worthwhile to note, however,
that the subtle long-range correlations, of the kind given in (\ref{corr gamma}),
although important in principle, might be not so prominent in practice.
For instance, according to the authors of \cite{Kondov}, their speckle
potential is well described by a (anisotropic) Gaussian correlation
function which is, by definition, \textquotedbl{}generic\textquotedbl{}
and short-ranged.

\subsection{$d=3$ and $\eta>>1$}

This is the case of a \textquotedbl{}smooth\textquotedbl{} random
potential (the condition $V_{0}>>E_{0}$ will be always satisfied
if the correlation length $R_{0}$ is made large enough). Since, in
contrast to the previous case, the energy $E_{0}$ is much below the
height of a typical barrier, the condition $kR_{0}>>1$ becomes now
the necessary condition for diffusive transport. We start with the
Born approximation, which should always hold at sufficiently high
energies. 

For the case $V_{0}>>E_{0}$, the Born approximation for a potential
barrier of strength $V_{0}$ and size $R_{0}$ results in a scattering
cross section \begin{eqnarray}
\sigma\sim\left(\frac{V_{0}}{kE_{0}}\right)^{2},\label{cross}\end{eqnarray}
 and the approximation is valid only if the particle energy satisfies
\cite{LL} \begin{eqnarray}
\epsilon>>\frac{V_{0}^{2}}{E_{0}}\equiv E_{\Delta}.\label{E}\end{eqnarray}
The scattering is strongly anisotropic and the transport cross section
$\sigma^{*}$ is smaller than $\sigma$ by a factor $(1/kR_{0})^{2}$,
which yields a transport mean free path given in Eq. (\ref{l'}).
The latter contains only quantities with a clear classical meaning,
and it can be arrived at by purely classical considerations. 
 Consider a particle with an initial velocity $v$, in a given direction.
The particle is subjected to a force of magnitude $|\textbf{F}|=|\nabla V(\textbf{r}|\sim V_{0}/R_{0}$.
The direction of the force is changing randomly, at a distance of
order $R_{0}$. During a time interval $\triangle t=R_{0}/v$ the
particle will acquire a velocity increment $|\triangle v|\sim F\triangle t/m\sim V_{0}/mv$.
This increment has a random direction, so that the number of time
intervals $\triangle t$, needed to completely degrade the initial
direction, is of order $(v/\triangle v)^{2}\sim(\epsilon/V_{0})^{2}$.
Thus, the time after which the initial direction is forgotten (the
transport time) is \begin{eqnarray}
\tau^{*}\sim\left(\frac{v}{\triangle v}\right)^{2}\triangle t\sim\left(\frac{\epsilon}{V_{0}}\right)^{2}\frac{R_{0}}{v},\label{tau}\end{eqnarray}
 and the corresponding mean free path is \begin{eqnarray}
l^{*}\sim v\tau^{*}\sim R_{0}\left(\frac{\epsilon}{V_{0}}\right)^{2}\;\;,\;\;\;\;\;\;\;\;\;(\epsilon>>V_{0})\label{l1}\end{eqnarray}
 which coincides with (\ref{l'}). This coincidence was pointed out
in \cite{DK} and it is quite remarkable, because Born approximation
is incompatible with semiclassics \cite{LL}. Indeed, the Born scattering
cross section, Eq. (\ref{cross}), strongly differs from the semiclassical
one, which is of order $R_{0}^{2}$. Also the differential cross sections
are entirely different: the typical scattering angle in the Born approximation
is $1/kR_{0}$, whereas in semiclassics it is $\triangle v/v\sim V_{0}/\epsilon$.
It is only the \textit{transport} scattering cross section (and hence
the transport mean free path) that is given by a similar expression
in both approaches. Furthermore, for $\epsilon<E_{\Delta}$ the Born
approximation breaks down. The estimate for $l^{*}$ in Eq. (\ref{l1}),
however, holds as long as $\epsilon>>V_{0}$. 
 Thus, we arrive at the conclusion that the only condition required
for a diffusive transport (with (\ref{l1}) serving as an order-of-magnitude
estimate for $l^{*}$) is $\epsilon>>V_{0}$, i.e. the particle energy
must well exceed the typical hight of the potential barriers. The
corresponding diffusion coefficient is \begin{eqnarray}
D_{\epsilon}=vl^{*}/3\sim D_{0}(\epsilon/V_{0})^{5/2},\label{DD}\end{eqnarray}
 where $D_{0}=(V_{0}R_{0}^{2}/m)^{1/2}$ is the natural unit for classical
diffusion (this is the only combination, of the three quantities involved,
with dimensions of a diffusion coefficient). 

When $\epsilon$ approaches $V_{0}$, the transport mean free path
approaches a value of order $R_{0}$. The energy interval
$\epsilon\sim V_{0}$ marks a transition from a
\textquotedbl{}soft\textquotedbl{}, small angle scattering to
\textquotedbl{}hard\textquotedbl{} scattering, when propagation
direction changes drastically in a single scattering event (in
this case there is no need to distinguish between $l$ and
$l^{*}$). When the particle energy becomes smaller than $V_{0}$,
and it is lowered towards the classical percolation threshold,
reflection from potential barriers starts to play an important
role and serves as \textquotedbl{}bottleneck\textquotedbl{} for
transport. The transport slows down, because the particle can get
temporarily trapped in a small region and it has to make many
bounces between the potential barriers before it escapes.
Eventually, at some critical energy $E_{c}$ transport (at zero
temperature) comes to complete halt. Classically, such transition
to an insulating state would have occurred at the percolation
threshold $E_{per}$, with a diffusion coefficient \cite{foot1}
behaving in the threshold vicinity as $D\sim
D_{0}[(\epsilon/E_{per})-1]^{t}$, where $t$ can be identified with
the percolation conductivity exponent,
close to $1.7$ in $3d$ \cite{ES,Zal}. 
Classical treatment, however, is never adequate close to the transition,
where quantum interference effects dominate. Thus, when the energy
is lowered towards $E_{per}$, the diffusion coefficient, at first,
will behave in accordance with the classical percolation theory but,
eventually, a crossover to the quantum (Anderson) transition must
take place \cite{Per} . This transition occurs at the mobility edge
$E_{c}$, which is strictly larger than $E_{per}$. (We stress that
this consideration applies only if $V_{0}>>E_{0}$, whereas in the
opposite case, considered in the previous subsection, no trace of
classical percolation appears in the transport picture.) The difference
$(E_{c}-E_{per})$ can be roughly estimated by locating the energy
at which the classical expression $D_{0}[(\epsilon/E_{per})-1]^{t}$
gets equal to the \textquotedbl{}diffusion quantum\textquotedbl{}
$\hbar/m$. Such estimate yields $E_{c}=E_{per}(1+\eta^{-1/2t})$.

One can also appreciate the importance of quantum effects, while moving
towards $E_{per}$ from the low energy side. Well below $E_{per}$
states are localized, essentially in a single potential well. When
$\epsilon$ is raised, a state spreads over several wells. At $\epsilon=E_{per}$,
a classical particle would get delocalized but quantum interference
prevent delocalization. Only when $\epsilon>E_{c}>E_{per}$, does
delocalization become possible.

The above relation between $E_{c}$ and $E_{per}$ is based on the
assumption of two well separated scales: a \textquotedbl{}microscopic\textquotedbl{}
scale, at which local diffusion (with a diffusion constant $D_{0}$)
takes place, and a macroscopic one, related to the topology of the
percolating cluster and responsible for the \textquotedbl{}slowing-down
factor\textquotedbl{} $[(\epsilon/E_{per})-1]^{t}$. This assumption
is by no means obvious even for the \textquotedbl{}generic\textquotedbl{}
random potentials, usually assumed in the percolation theory \cite{ES,Zal}.
Moreover, one can construct \textquotedbl{}unusual\textquotedbl{}
potentials, the extreme example being a potential which, although
random, is strictly zero within a connected spatial region. Percolation
is then possible at all energies, and there is no connection at all
between classical percolation and quantum localization. Actually,
a blue detuned speckle might be not so far from this extreme example
since, due to the high probability for low values of the potential,
the percolation threshold (in $3d$) is only about $4\cdot10^{-4}V_{0}$
\cite{Pil}.

\subsection{$d=2$ and $\eta<<1$}

In two dimensions all states are localized. If, however, the
disorder (at the given energy) is weak, i.e. $kl>>1$, the
localization length is exponentially large. In this case, in a
finite size sample, the wave function can easily spread over the
entire sample, and the states can be considered as extended, for
all practical purposes. When disordered increases (or the particle
energy decreases, for fixed disorder), the localization length
becomes smaller and, eventually, the regime of strong localization
is reached. Thus, in $2d$, one speaks about a crossover from weak
to strong localization, instead of a strict transition between
extended and localized states that is taking place in $3d$.

This picture can be rephrased in terms of transport. At high energies
the ordinary diffusive transport is possible. Indeed, it would take
unrealistically large samples and exponentially long times (in addition
to the almost complete absence of any inelastic processes) to find
out that the initial wave packet gets, in fact, localized. Under appropriate
conditions, one can observe weak localization corrections, on top
of diffusion \cite{Ab,Lee,Akker}. When the energy is lowered, the
diffusion coefficient decreases and one crosses over to the strong
localization regime: the localization length becomes much smaller
than the sample size and transport is inhibited.

As in Sec. $3.1$, we start with an estimation of the cross section
for a single scatterer- a potential barrier of hight $V_{0}$ and
radius $R_{0}$. For a slow particle, $kR_{0}<<1$, this cross section
in the Born approximation is \cite{LL} $\sigma\sim\eta^{2}/k$ and,
unlike the $3d$ case, it is energy dependent. The approximation is
valid for $kR_{0}>>\eta^{2}$. Next, we estimate the mean free path
for a particle moving in the presence of many scatterers, in concentration
$n_{scat}\sim R_{0}^{-2}$: \begin{eqnarray}
l=\frac{R_{0}^{2}}{\sigma}\sim\frac{kR_{0}^{2}}{\eta^{2}}.\label{mfp2}\end{eqnarray}
 Furthermore, the Ioffe-Regel criterion, $kl>>1$, results in the
condition $kR_{0}>>\eta$ (this is more restrictive than the condition
$kR_{0}>>\eta^{2}$ for a single scatterer). In terms of the particle
energy one obtains \cite{Mini,Raikh}: \begin{eqnarray}
\epsilon>>V_{0}\eta\equiv E_{\Delta}.\label{E2}\end{eqnarray}
 This energy scale has already appeared in (\ref{E}), although its
significance here is quite different from that in Sec. $3.2$, devoted
to the smooth $3d$ potential.

Consider now the fast particles, $kR_{0}>>1$. The transport
scattering cross section for those differs from
$\sigma\sim\eta^{2}/k$ by a product of two small factors (compare
to the corresponding estimate in Sec. $3.1$). The factor
$1/kR_{0}$ accounts for the narrow scattering sector, in the
forward direction, and the factor $1/(kR_{0})^{2}$ accounts for
the difference between the scattering and the transport cross
sections. Thus, $\sigma^{*}\sim\eta^{2}/k^{4}R_{0}^{3}$, and the
transport mean free path $l^{*}=R_{0}^{2}/\sigma^{*}\sim
R_{0}(\epsilon/V_{0})^{2}$, as in (\ref{l'}).

The quantitative treatment of $l^{*}$, in the Born approximation,
is based on the $2d$ counterpart of (\ref{inverse}):
\begin{eqnarray}
\frac{1}{\tau^{*}}=\frac{1}{\hbar}V_{0}^{2}\nu\int_{0}^{2\pi}d\phi\tilde{\Gamma}
\left(2k\sin\frac{\phi}{2}\right)(1-\cos\phi),\label{inverse2}\end{eqnarray}
 where $\nu=m/2\pi\hbar^{2}$ is the $2d$ density of states and $\tilde{\Gamma}$
is the Fourier transform of the correlation function
$\Gamma(R/R_{0})$ for the $2d$ random potential. For a Gaussian
correlation function, $\Gamma=\exp(-R^{2}/R_{0}^{2}$), the result
for $l^{*}=\hbar k\tau^{*}/m$ is \cite{Mini,Raikh,BS1}:
\begin{eqnarray}
l^{*}(\epsilon)=\frac{R_{0}}{\pi}\left(\frac{\epsilon}{E_{0}}\right)^{1/2}\left(\frac{2E_{0}}{V_{0}}\right)^{2}f\left(\frac{\epsilon}{2E_{0}}\right)\;\;\;\;\;,\;\;\;\;(\epsilon>>E_{\Delta}).\label{l''2}\end{eqnarray}
 Here the function $f(x)=e^{x}[I_{0}(x)-I_{1}(x)]^{-1}$, where $I_{0}$
and $I_{1}$ are the modified Bessel functions. For small $x$, $f(x)\approx1$;
for large $x$, $f(x)\approx\sqrt{\pi}(2x)^{3/2}$. In these limits
of slow and fast particles, respectively, one can immediately write
down the corresponding values for the mean free path, in complete
agreement with the above estimates. In terms of the diffusion coefficient,
$D_{\epsilon}=vl^{*}/2$, the limiting values are: \begin{equation}
D_{\epsilon}=\left\{ \begin{array}{c}
\frac{2\hbar}{\pi m}\left(\frac{\epsilon-\overline{V}}{E_{\Delta}}\right)\;\;\;\;\;,\;\;\;\;\;\; E_{\Delta}<<(\epsilon-\overline{V})<<E_{0}\\
(\frac{8V_{0}R_{0}^{2}}{\pi m})^{1/2}\left(\frac{\epsilon-\overline{V}}{V_{0}}\right)^{5/2}\;\;,\;\;\;\;\;\;\;\;\;\epsilon>>E_{0}\end{array}\right.\label{D2}\end{equation}
 where, as in (\ref{D}), we have made it explicit that the energy
should be counted from the average potential value.

The important case of a correlation function, corresponding to a speckle
potential created by a uniformly illuminated circular diffusive plate,
was considered in \cite{Mini}. Although that correlation function
decays only as $R^{-3}$ (with oscillations), the integral $\int\Gamma(\textbf{R})d^{2}R$
is finite, so that this case is, qualitatively, not different from
the Gaussian correlation. The equation (\ref{D2}) still holds, albeit
with different numerical coefficients (note that our definition of
$E_{0}$ and $\eta$ differs by a factor of $2$ from that in \cite{Mini}).

Unlike the $3d$ case, when the condition $(\epsilon-\overline{V})>>W$
was sufficient for diffusive transport, in $2d$ the condition $(\epsilon-\overline{V})>>E_{\Delta}$
is only a necessary one. In an ideal world of infinite samples, infinite
times and zero temperature, quantum interference effects would eventually
take over and transport would cease. In reality, factors like dephasing
or finite sample size suppress interference and provide room for diffusion.
Interference effects show up in a correction to the diffusion coefficient
(weak localization). As the just mentioned \textquotedbl{}ideal conditions\textquotedbl{}
are approached, interference becomes more important and the diffusion
coefficient approaches zero. All this is discussed in great detail
in \cite{Kuhn,Mini}. For $(\epsilon-\overline{V})\sim E_{\Delta}$
a crossover to strong localization takes place. The localization length
$\xi$ rapidly decreases with energy, and for $(\overline{V}-\epsilon)>>E_{\Delta}$
the region of strongly localized states, well separated from each
other, is reached. This region of the spectrum has been already discussed
in Sec. $3.1$, one should only set $d=2$ in the corresponding expressions.

In condensed matter physics, transport in the region of strongly localized
states can occur only at finite temperatures, and it is due to hopping
\cite{ES,Pollak}: an electron can hop from one localized state to
another by absorbing a phonon. In principle, one should be able to
observe a similar phenomenon for localized cold atoms, by modelling
phonons with the help of a random, time-dependent optical potential
(this remark applies, of course, also to $3d$).

\subsection{$d=2$ and $\eta>>1$}

A few brief remarks will suffice to summarize the situation for this
case. Using considerations entirely similar to those in Sec. $3.2$,
one can see that Eq. (\ref{l1}) for the transport mean free path
is valid also in $2d$, as long as $\epsilon>>V_{0}$. 
The corresponding diffusion coefficient is $D=vl^{*}/2\sim D_{0}(\epsilon/V_{0})^{5/2}$,
i.e. the same as in $3d$, up to a numerical factor.

When $\epsilon$ approaches the typical barrier height $V_{0}$, the
diffusion coefficient reaches a value $\sim D_{0}$, and it keeps
decreasing when the energy is lowered further. The decrease occurs
for two reasons: first, it becomes more a more difficult for a (classical)
particle to find a path around the potential barriers and, second,
the quantum interference effects are being enhanced, leading to strong
localization. When $D$ reaches a value of order $\hbar/m$, the classical
picture breaks down completely. Diffusion of a classical particle
in a $2d$ speckle potential has been numerically studied in \cite{Pezze},
and it would be interesting to account for quantum effects, including
the classical-quantum crossover.

\section{A superfluid-insulator transition in a disordered BEC}

The behavior of a BEC in a random environment is a vast subject, with
a long history. The question of how disorder can affect, and possibly
destroy, superfluidity and superconductivity has been repeatedly discussed
in the literature \cite{Rep,Lee}. In particular, Ref. \cite{Rep}
contains a thorough discussion on $^{4}He$ absorbed in a porous Vycor
glass. The porous, random structure of the medium can lead to a significant
reduction of the superfluid transition temperature $T_{c}$.

Ultracold atomic gases, which allow for a very accurate control over
both the disorder and the interparticle interaction, provide an ideal
system for studying the interplay between the two factors. The problem
is of fundamental interest, because it brings together two central
themes in condensed matter physics: interactions and disorder. In
the absence of interactions all bosons would drop into the lowest
potential well, so the ideal Bose gas is clearly not the appropriate
starting point for the theory. Any finite repulsive interaction has
a drastic effect, pushing the particles apart and not allowing for
too high local density $n(\textbf{r})$. If the average particle density,
$n$, is finite, the particles will get spread over the entire (infinite)
system.

We will assume the model of a continuous random potential
V(\textbf{r}) (see previous section), thus, leaving out the work
on disordered (or quasiperiodic) optical lattices and the
disordered Bose-Hubbard model (see \cite{Lew} for an extensive
discussion and references,  and \cite{Pas,Dei} for recent
experiments). Furthermore, we will not discuss finite temperature
effects or excitations above ground state \cite{Pil,Gau,Lug, Al1}
but focus on the zero-temperature, quantum phase transition, i.e.,
the transition between a superfluid and an insulator under a
change of the chemical potential $\mu$ (or the particle density
$n$) \cite{Shk,Fal}.
Ref. \cite{Fal} contains a detailed exposition of various regimes,
in $2d$ and $3d$, including the effect of a harmonic trap. We refer
the reader to \cite{Fal} for details, and here only review the transition
in $3d$, for a short range random potential, $\eta<<1$.

We consider, thus, $N$ bosons in a box of volume $\Omega$, in the
thermodynamic limit ($N,\Omega\rightarrow\infty,n=N/\Omega=const$).
Repulsive interactions and the \textquotedbl{}gas condition\textquotedbl{},
$na^{3}<<1$, are assumed. Here $a$ is the scattering length, related
to the interaction constant as $g=4\pi\hbar^{2}a/m$. In the absence
of disorder, the chemical potential is $\mu=gn$ \cite{String,Pet}.
A weak random potential can be treated in perturbation theory \cite{Huang,Gior}.
The important energy, characterizing a short range random potential,
is $W$ (see Eq.(\ref{W})). A particle with the de-Broglie wave length
$\lambda_{p}>>R_{0}$ experiences not the \textquotedbl{}bare\textquotedbl{}
potential $V({\bf r)}$ but a smoothed, effective potential, and the
scattering can be considered weak as long as $\lambda_{p}<<l\sim R_{0}/\eta^{2}$.
For a BEC the role of $\lambda_{p}$ is played by the healing length
$\xi_{h}=\hbar/\sqrt{2mgn}=\hbar/\sqrt{2m\mu}$ and the effect of
the potential is small, provided the condition $\xi_{h}<<l$, i.e.
$\mu>>W$, is satisfied. Thus, for $\mu>>W$ one has a weakly disordered
BEC. It is described by a macroscopic wave function $\Psi({\bf {r})}$
which, although fluctuating in space, maintains coherence over the
entire box. The system is a superfluid.

The perturbation theory is expected to break down when $\mu$ reaches
a value of order $W$, i.e. when particle concentration becomes of
order $W/g\sim1/\mathcal{L}^{2}a$, where we have introduced the Larkin
length defined in Sec. 3.1. At such concentrations disorder becomes
more important, fluctuations grow stronger and at some point the coherence
gets disrupted: the superfluidity is destroyed and the system undergoes
a transition to the insulating state. Let's denote this critical concentration
by $n_{c}$ and the corresponding value of the chemical potential
$\mu=\mu_{c}$. Generally, there is no simple relation between $\mu_{c}$
and the mobility edge $E_{c}$ (or the energy $W$). To determine
$\mu_{c}$ one has to solve the many body problem of interacting disordered
bosons, whereas $E_{c}$ is a single particle property. In other words,
there is no straightforward connection between superfluidity and the
nature of the single-particle states (localized vs extended). Indeed,
it has been argued long ago \cite{DKKL} that, due to the interaction
induced screening, a condensate can exist when the chemical potential
is well below $E_{c}$. 
 Nevertheless, an estimate of $\mu_{c}$ for weakly interacting disordered
bosons \cite{Shk,Fal} shows that $W$ is an important energy scale
in the problem. 
The arguments and estimates in \cite{Shk,Fal} were made for a
Gaussian random potential. Since, however, the main
\textquotedbl{}action\textquotedbl{} is taking place in a narrow
interval near $\overline{V}$, the considerations should apply also
to a speckle potential. For small $n$, when $\mu$ (measured from
the average potential value $\overline{V}$) is negative and large
in the absolute value, the distribution of particles in space is
highly inhomogeneous. The particles, in the ground state, fill the
potential wells with radius smaller than some value $R$ (the
corresponding depth is $\sim-\hbar^{2}/mR^{2}$), and the purpose
is to determine this value, for a given $n$. Let's denote by
$n_{w}(R)$ the concentration of such wells. For $R$ small, namely
$R<<\mathcal{L}$, the typical distance between neighboring wells
is exponentially large and their concentration is \cite{Fal}
\begin{eqnarray}
n_{w}(R)=\frac{1}{R^{3}}f\left(\frac{\mathcal{L}}{R}\right)\exp\left(\frac{-\mathcal{L}}{R}\right),\label{nw}\end{eqnarray}
 where the preexponent $f$ plays no significant role in the forthcoming
estimate. The average number of particles per well is $N_{w}(R)=n/n_{w}(R)$
and the particle density in a well is $n_{p}(R)=3N_{w}(R)/4\pi R^{3}$.
 The total energy of $N_{w}(R)$ bosons in such a well is comprised
of two parts: the binding energy and the interaction energy. Introducing
the energy per particle, $E(R)$, one can write (with somewhat arbitrary
numerical coefficients):
\begin{eqnarray}
E(R)=-\frac{\hbar^{2}}{2mR^{2}}+gn_{p}(R)=-\frac{\hbar^{2}}{2mR^{2}}+\frac{3\hbar^{2}na}{n_{w}(R)mR^{3}}\;.
\label{Etot}\end{eqnarray}
 Now, using (\ref{nw}) and minimizing (\ref{Etot}) with respect
to $R$ (for fixed $n$), one obtains \cite{Fal}: \begin{eqnarray}
R(n)=\frac{\mathcal{L}}{ln(n_{c}/n)}\;\;,\;\;\;\;\;\;\;\;\;\;
n_{c}=\frac{1}{3\mathcal{L}^{2}a}\;.\label{Rn}\end{eqnarray}
 For this result to be valid, one must require $n<<n_{c}$. On the
other hand, since the number $N_{w}(R)$ must be large, $n$ should
be not too small, namely: \begin{eqnarray}
n_{c}\exp\left[-(\mathcal{L}/3a)^{1/3}\right]<<n<<n_{c},\label{ineq}\end{eqnarray}
 which is consistent only if $\mathcal{L}>>a$. Under the condition
(\ref{ineq}), tunneling between the wells is negligible and the
number of particles in each well is fixed. The state is insulating
(a fragmented condensate). The chemical potential is estimated as
\begin{eqnarray}
\mu(n)\sim-\frac{\hbar^{2}}{mR^{2}(n)}\sim-W\ln^{2}\left(\frac{n_{c}}{n}\right).\label{mu}\end{eqnarray}

When $n$ increases, the size $R(n)$ of the relevant wells gets larger
and, for $n\sim n_{c}$, it becomes comparable to the separation between
the wells (both are of order $\mathcal{L}$). Tunneling becomes significant
and at some point a transition to the coherent superfluid state occurs.
Of course, $n_{c}$, as defined in (\ref{Rn}), is not the precise
transition point but only an order of magnitude estimate. Similarly,
the critical value $\mu_{c}$ of the chemical potential is not accurately
known. 


\section{Dynamics of cold atoms in the presence of disorder}

The present section is devoted to the time evolution of atomic clouds
subjected to a random potential. In a typical experimental set-up
a ultracold atomic gas, of either bosons or fermions, is released
from a harmonic trap, $V_{tr}(r)=m\omega^{2}r^{2}/2$, and undergoes
expansion while being scattered by a random potential $V({\bf r)}$.
The trap determines the initial state of the gas, whereas the potential
$V({\bf r)}$ affects its subsequent time evolution. At some moment
$t$ an image of the atomic cloud is taken, providing information
about the mode of transport. We discuss separately two atomic systems:
BEC and a cold Fermi gas.

\subsection{Ballistic expansion of a BEC}

We start with the problem of a free, ballistic expansion, i.e. the
random potential is absent. BEC in the trap is described by a macroscopic
wave function, satisfying the stationary Gross-Pitaevskii equation
\begin{equation}
-\frac{\hbar^{2}}{2m}\triangle\Psi+\frac{1}{2}m\omega^{2}r^{2}\Psi+g|\Psi|^{2}\Psi=\mu\Psi.\label{G1}\end{equation}
 Usually, interaction among the atoms in the trap is of great importance:
the size of the condensate, $R_{\mu}\sim(\mu/m\omega^{2})^{1/2}$,
is much larger than the \textquotedbl{}oscillator size\textquotedbl{}
$a_{0}=(\hbar/m\omega)^{1/2}$. The latter is, roughly, the size of
a single particle ground state wave function, and in the absence of
interactions all the bosons would have condensed into this single
particle state. The inequality $R_{\mu}>>a_{0}$ implies $\mu>>\hbar\omega$.
This is the condition for validity of the Thomas-Fermi approximation
which amounts to the neglect of the kinetic energy term in (\ref{G1}),
thus, resulting in a simple expression (an inverted parabola) for
the condensate density \begin{equation}
n(\textbf{r},t=0)=|\Psi(\textbf{r},t=0)|^{2}=\frac{\mu}{g}\left(1-\frac{r^{2}}{R_{\mu}^{2}}\right)\;\;,\;\;\;\;\;\;\;\;\;\; R_{\mu}=\left(\frac{2\mu}{m\omega^{2}}\right)^{1/2}\label{n1}\end{equation}
 where we have added the time argument $t=0$, to emphasize that this
is the condensate profile in the trap, prior to its release. The corresponding
wave function is a square root of this expression, with an arbitrary
overall phase which can be set to zero: \begin{equation}
\Psi(\textbf{r},t=0)=\left[\frac{\mu}{g}\left(1-\frac{r^{2}}{R_{\mu}^{2}}\right)\right]^{1/2}\equiv F(r).\label{Psi1}\end{equation}
 This expression is accurate in the bulk of the condensate, and it
breaks down only close to the boundary, for $r\approx R_{\mu}$.

Assume now that at $t=0$ the trap potential is switched off, and
for $t>0$ the BEC undergoes free evolution, according to the time
dependent Gross-Pitaevskii equation: \begin{equation}
i\hbar\frac{\partial\Psi}{\partial t}=-\frac{\hbar^{2}}{2m}\triangle\Psi+g|\Psi|^{2}\Psi\;\;,\;\;\;\;\;\;\;\;\;\;\int{d{\bf r}}|\Psi|^{2}=N.\label{G2}\end{equation}
 with the initial condition given in (\ref{Psi1}). The wave function
is normalized to the total number of particles $N$.
$\Psi(\textbf{r},t)$ is a complex function,
$\Psi=\sqrt{n}\exp(i\theta)$, i.e. the condensate acquires a
dynamic phase $\theta(\textbf{r},t)$. Eq. (\ref{G2}) can be
rewritten as a set of two hydrodynamic equations, in terms of the
condensate density $n(\textbf{r},t)$ and its velocity
$\textbf{v}(\textbf{r},t)=(\hbar/m)\nabla\theta(\textbf{r},t)$
\cite{String,Pet}: \begin{equation} \frac{\partial n}{\partial
t}+divn\textbf{v}=0\label{con}\end{equation}
 \begin{equation}
m\frac{\partial\textbf{v}}{\partial t}+\nabla\left(\frac{1}{2}mv^{2}+gn\right)=0,\label{euler}\end{equation}
 where in the second equation the \textquotedbl{}quantum pressure\textquotedbl{}
term has been neglected, which is justified if the healing length
$\xi_{h}$ is smaller than the characteristic scale over which the
density is changing (see the discussion in \cite{Pet} ). These equations
have to be solved with the initial conditions $\textbf{v}(\textbf{r},t=0)=0$
and $n(\textbf{r},t=0)$ given in (\ref{n1}). The solution is \cite{String,Kag,Cas}:
\begin{equation}
n(\textbf{r},t)=\frac{\mu}{gb^{d}}\left(1-\frac{r^{2}}{b^{2}R_{\mu}^{2}}\right),\;\;\;\;\;\;\;\;\;\;\;\;\textbf{v}(\textbf{r},t)=\frac{\dot{b}}{b}\textbf{r},\label{nv}\end{equation}
 where the scaling factor $b(t)$ obeys the ordinary differential
equation $\ddot{b}=\omega^{2}/b^{d-1}$, with $b(0)=1,\dot{b}(0)=0$
(a dot on $b$ denotes time derivative). At $d=2$ its solution is
$b(t)=\sqrt{1+\omega^{2}t^{2}}$, and at $d=3$ it is qualitatively
similar, namely: $b\approx1$ for $t<<1/\omega$ and $b\approx\omega t$
for $t>>1/\omega$ (linear evolution).

Thus, in the course of the expansion the condensate density
$n(\textbf{r},t)$ retains its shape of an inverted parabola whose
size increases according to the scaling factor $b(t)$. The
important time scale in the expansion process is $t_{0}=1/\omega$.
For $t<<t_{0}$ the time evolution is dominated by the nonlinear
term in (\ref{G2}) and the condensate energy is due primarily to
the interaction. By the time $t\sim t_{0}$ most of the interaction
energy gets converted into the kinetic energy of the condensate
flow. The wave function $\Psi(\textbf{r},t_{0})$ exhibits rapid
spatial oscillations which account for the large kinetic energy of
the condensate. For $t>>t_{0}$ the interaction term (i.e. the
nonlinear term in (\ref{G2})) becomes negligible and the wave
function evolves according to the linear Schrödinger equation. The
function at $t=t_{0}$ is, roughly, \begin{equation}
\Psi(\textbf{r},t_{0})\simeq
F(\frac{r}{2})\exp(ir^{2}/a_{0}^{2})\equiv\Phi(r),\label{Phi}\end{equation}
 where the argument $r/2$ in $F$ indicates that by the time $t_{0}$
the condensate's size has increased by a factor of 2 or so. One can
check that the kinetic energy in this wave packet, $E_{kin}=\frac{\hbar^{2}}{2m}\int|\nabla\Psi|^{2}d\textbf{r}$,
is of the same order of magnitude as the interaction energy, $E_{int}=\frac{g}{2}\int|\Psi|^{4}d\textbf{r}$.

\subsection{Diffusion of a BEC}

Here we study the time evolution of a BEC, in a random potential $V({\bf r)}$,
upon its release from the trap. The evolution is governed by the Gross-Pitaevskii
equation, with the random potential: \begin{equation}
i\hbar\frac{\partial\Psi}{\partial t}=-\frac{\hbar^{2}}{2m}\triangle\Psi+V({\bf r})\Psi+g|\Psi|^{2}\Psi\label{G3}\end{equation}
 Sometimes it is possible to disentangle the effects of nonlinearity
and disorder \cite{SP2,BS1}. For instance, one can switch the trap
off at $t=0$, let the condensate freely expand for a time interval
of the order of $t_{0}=1/\omega$, and only then switch on the
random potential. In this way the evolution is separated into two
stages. The first stage, for times $t\sim t_{0}$, is dominated by
the nonlinearity (there is no disorder yet), whereas at the second
stage, $t>>t_{0}$, the nonlinearity can be neglected and the
evolution proceeds according to the Schrödinger equation
\begin{equation} i\hbar\frac{\partial\Psi}{\partial
t}=-\frac{\hbar^{2}}{2m}\triangle\Psi+V({\bf
r})\Psi\;\;,\;\;\;\;\;\;\;\;\;\;(t>t_{0}),\label{Sch1}\end{equation}
 with the initial condition $\Psi(\textbf{r},t=t_{0})=\Phi(r)$, see
(\ref{Phi}). Setting the start of the second stage at $t=t_{0}$
is, of course, only a rough estimate (one should wait for a time few
times larger than $t_{0}$ before switching on $V({\bf r)}$) but
it does capture the main idea of the two-stage evolution.

In a different set-up, one releases the condensate, at $t=0$, directly
into a random potential (rather than switching on the potential later)
and the time evolution for $t>0$ follows the equation (\ref{G3}).
If $V({\bf r)}$ is sufficiently weak, one can still argue in favor
of a two-stage scenario. During the time $t\sim t_{0}$ nonlinearity
dominates over disorder and the expansion is essentially ballistic
(Sec. 5.1). For $t>>t_{0}$ the nonlinearity is weak and disorder
becomes the important factor: the evolution is close to linear, according
to (\ref{Sch1}). Whether a given potential $V({\bf r)}$ is weak
or strong depends on the wave number k of the relevant component of
the condensate wave function. If, for instance, the chemical potential
$\mu$ in the trap is much larger than the mobility edge $E_{c}$,
then most of the atoms, upon their release from the trap, will experience
weak disorder, because the parameter $k_{\mu}l$ is large ($k_{\mu}=\sqrt{2m\mu/\hbar^{2}}$).
On the other hand, spreading of the small-$k$ components might be
inhibited by disorder and they will stay localized in the vicinity
of the trap. For these components the nonlinearity might remain important
for all times. In any case, the first set-up, i.e., when the potential
is switched on after sufficiently long interval of a ballistic expansion,
is better suited for the two-stage scenario.

Assuming the validity of this scenario, it remains to study the linear
evolution (\ref{Sch1}), with the initial condition (\ref{Phi}).
This condition carries information about the initial state of the
condensate in the trap and about the first stage of expansion, dominated
by the nonlinearity. To simplify the notations, we reset the time
$t_{0}$ to zero, so that the formal solution of (\ref{Sch1}) is
\begin{equation}
\Psi(\textbf{r},t)=\int d\textbf{R}G(\textbf{r},\textbf{R},t)\Phi(\textbf{R}),\label{Psi3}\end{equation}
 where $G$ is the retarded Green's function of the Schrödinger equation
(\ref{Sch1}). The average particle density \begin{equation}
\overline{n}(\textbf{r},t)=\overline{|\Psi(\textbf{r},t)|^{2}}=\int d\textbf{R}\int d\textbf{R}'\overline{G^{*}(\textbf{r},\textbf{R},t)G(\textbf{r},\textbf{R}',t)}\Phi^{*}(\textbf{R})\Phi(\textbf{R}').\label{n av}\end{equation}
 The product of the Green's functions can be Fourier transformed as
\begin{equation}
\overline{G^{*}(\textbf{r},\textbf{R},t)G(\textbf{r},\textbf{R}',t)}=\int\frac{d\varepsilon}{2\pi}\int\frac{d\Omega}{2\pi}e^{-\frac{i\Omega t}{\hbar}}\overline{G^{*}(\textbf{r},\textbf{R},\varepsilon+\frac{1}{2}\Omega)G(\textbf{r},\textbf{R}',\varepsilon-\frac{1}{2}\Omega)}\:.\label{GG}\end{equation}
 The product in (\ref{GG}), for weak disorder, can be calculated
in the diagram technique \cite{Akker}: \begin{equation}
\overline{G^{*}(\textbf{r},\textbf{R},\varepsilon+\frac{1}{2}\Omega)G(\textbf{r},\textbf{R}',\varepsilon-\frac{1}{2}\Omega)}=-2P_{\varepsilon}(\textbf{r},\textbf{R},\Omega)Im\overline{G}(\textbf{R}-\textbf{R}',\epsilon),\label{ladder}\end{equation}
 where the diffusion ladder is defined as \begin{equation}
P_{\varepsilon}(\textbf{r},\textbf{R},\Omega)=\frac{1}{2\pi\nu_{\epsilon}}\overline{G^{*}(\textbf{r},\textbf{R},\varepsilon+\frac{1}{2}\Omega)G(\textbf{r},\textbf{R},\varepsilon-\frac{1}{2}\Omega)}\end{equation}
 and the average Green's function \begin{equation}
\overline{G}(\boldsymbol{\rho},\epsilon)=G_{0}(\boldsymbol{\rho},\varepsilon)e^{-\frac{\rho}{2l_{\varepsilon}}},\;\;\;\;\;\;\;\;\;\;\;\;(\boldsymbol{\rho}=\textbf{R}-\textbf{R}').\label{the average Green function}\end{equation}
 Here $G_{0}$ is the free Green's function and $\nu_{\epsilon}$
and $l_{\epsilon}$ are, respectively, the (average) density of states
and the mean free path. Putting all the pieces together, we arrive
at the following expression for the particle density: \begin{equation}
\overline{n}(\textbf{r},t)=-\frac{1}{\pi}\int d\textbf{R}\int d\boldsymbol{\rho}\int d\epsilon P_{\epsilon}(\textbf{r},\textbf{R},t)Im\overline{G}(\boldsymbol{\rho},\epsilon)\Phi^{*}(\textbf{R}+\frac{\boldsymbol{\rho}}{2})\Phi(\textbf{R}-\frac{\boldsymbol{\rho}}{2}),\label{n1 av}\end{equation}
 where the quantum diffusion kernel $P_{\epsilon}(\textbf{r},\textbf{R},t)$
is the Fourier transform of the diffusion ladder $P_{\varepsilon}(\textbf{r},\textbf{R},\Omega)$.
Although this result was derived for weak disorder, scaling arguments
(as presented e.g. in \cite{Lee,Chalk}) suggest that, with the appropriate
form for the kernel $P$, the result is valid also when disorder is
strong. Transforming $\overline{G}(\boldsymbol{\rho},\epsilon)$ to
the momentum representation, $\overline{G}(\textbf{k},\epsilon)$,
enables us to write \begin{equation}
\overline{n}(\textbf{r},t)=-\frac{1}{\pi}\int d\textbf{R}\int d\textbf{k}\int d\epsilon P_{\epsilon}(\textbf{r},\textbf{R},t)Im\overline{G}(\textbf{k},\epsilon)W(\textbf{k},\textbf{R}),\label{n2 av}\end{equation}
 where the Wigner function is defined as \begin{equation}
W(\textbf{k},\textbf{R})=\frac{1}{(2\pi)^{d}}\int d\boldsymbol{\rho}e^{i\textbf{k}\cdot\boldsymbol{\rho}}\Phi^{*}(\textbf{R}+\frac{\boldsymbol{\rho}}{2})\Phi(\textbf{R}-\frac{\boldsymbol{\rho}}{2}).\label{Wigner}\end{equation}

Eq. (\ref{n2 av}) contains three ingredients: (i)The Wigner function
carries information about the initial condition. (ii)The factor $-(1/\pi)Im\overline{G}(\textbf{k},\epsilon)\equiv A(\textbf{k},\epsilon)$
is the spectral function. For a free particle it is equal to $\delta(\epsilon-\epsilon_{\textbf{k}})$,
with $\epsilon_{\textbf{k}}=\hbar^{2}k^{2}/2m$ (the \textquotedbl{}on-shell\textquotedbl{}
relation between the energy and momentum). Disorder broadens the $\delta$-function
into a Lorentzian or, if sufficiently strong, into some more complicated
shape. (iii) The quantum diffusion kernel propagates a particle, with
energy $\epsilon$, from $\textbf{R}$ to $\textbf{r}$, in time $t$.

Eq. (\ref{n2 av}) simplifies in certain limits. If one uses $\delta(\epsilon-\epsilon_{\textbf{k}})$
for the spectral function, which can be justified for sufficiently
weak disorder, one obtains \begin{equation}
\overline{n}(\textbf{r},t)=\int d\textbf{R}\int d\textbf{k}P_{k}(\textbf{r},\textbf{R},t)W(\textbf{k},\textbf{R}),\label{App1}\end{equation}
 where $P_{k}\equiv P_{\epsilon_{k}}$. This equation appears, in
a somewhat different form, in \cite{Mini} and it has a simple interpretation:
The initial distribution in phase space, $W(\textbf{k},\textbf{R})$,
is propagated in time by the propagation kernel $P_{k}(\textbf{r},\textbf{R},t)$.
One can rewrite this equation in terms of the momentum $\textbf{p}=\hbar\textbf{k}$,
making it look entirely classical. Note, however, that both $W$ and
$P_{k}$ are quantum objects.

Another type of approximation, appropriate for large distances $r$,
is based on neglecting the $R$-dependence of $P_{k}(\textbf{r},\textbf{R},t)$.
For particles which had propagated a distance $r>>R_{\mu}$, the trap
(i.e. the initial size of the cloud $R_{\mu})$ can be viewed as a
\textquotedbl{}point object\textquotedbl{} and $R$ can be set equal
to zero (the center of the trap). The integral of $W(\textbf{k},\textbf{R})$
over $\textbf{R}$ gives the momentum distribution $|\tilde{\phi}(\textbf{k})|^{2}$,
where $\tilde{\phi}(\textbf{k})$ is the Fourier transform of the
initial wave function $\Phi(\textbf{r})$. (\ref{n2 av}) now reduces
to \cite{Skip} 
\begin{equation}
\overline{n}(\textbf{r},t)=\int\frac{d\textbf{k}}{(2\pi)^{d}}|\tilde{\phi}(\textbf{k}|^{2}\int d\epsilon P_{\epsilon}(\textbf{r},0,t)A(\textbf{k},\epsilon).\label{App2}\end{equation}
 A still simpler expression for $\overline{n}(\textbf{r},t)$ is obtained
by combining the two aforementioned approximation, i.e. by equating
the spectral function in (\ref{App2}) with $\delta(\epsilon-\epsilon_{\textbf{k}})$:
\begin{equation}
\overline{n}(\textbf{r},t)=\int\frac{d\textbf{k}}{(2\pi)^{d}}|\tilde{\phi}(\textbf{k}|^{2}P_{k}(\textbf{r},t),\label{App3}\end{equation}
 where the second argument, $\textbf{R}=0$, in $P_{k}(\textbf{r},\textbf{R},t)$
is omitted. The expression (\ref{App3}) can be visualized as dynamical
evolution of a cloud of classical particles, which are initially at
the origin and whose velocity distribution is determined by $|\tilde{\phi}(\textbf{k})|^{2}$.
Thus, (\ref{App3}) applies either to a BEC (a single coherent wave
packet) or to a cloud of independent, non-interacting particles, e.g.,
a Fermi gas (see Sec. 5.4). The information about the system is in
the initial distribution $|\tilde{\phi}(\textbf{k})|^{2}$, 
which can be, for instance, the thermal distribution (Bose or Fermi)
at a given temperature.

The interpretation of $\overline{|\Psi(\textbf{r},t)|^{2}}$ in terms
of \textquotedbl{}particles\textquotedbl{} is not limited to (\ref{App3})
but holds generally. One can imagine performing integration over $\textbf{R}$
and $\textbf{k}$ in (\ref{n2 av}), ending up with the spectral decomposition
of the condensate density $\overline{n}(\textbf{r},t)=\int\frac{d\epsilon}{2\pi}\overline{n}_{\epsilon}(\textbf{r},t)$.
Similarly, integration over $\textbf{R}$ and $\epsilon$ would produce
the momentum decomposition $\int\frac{d\textbf{k}}{(2\pi)^{d}}\overline{n}_{\textbf{k}}(\textbf{r},t)$.
Thus, one can interpret the total density $\overline{n}(\textbf{r},t)$
as being composed of fictitious particles, or atoms, with a given
energy (or momentum), and in what follows we often use this intuitive
picture, talking for instance about fast or slow particles. One should
keep in mind, though, that the condensate wave function is a coherent
unit and that only the total density has the straightforward interpretation
in terms of the actual bosonic atoms.

To proceed further one needs an expression for the quantum diffusion
kernel $P_{\epsilon}(\textbf{r},\textbf{R},t)$. This is the main
object of the theory because it determines the dynamics, for a given
$\epsilon$. For weak disorder, when propagation is by simple diffusion,
$P$ is the standard diffusion propagator: \begin{equation}
P_{\varepsilon}(\textbf{r},\textbf{R},t)=\frac{1}{\left(4\pi D_{\varepsilon}t\right)^{d/2}}\exp\left(-\frac{\left|\textbf{r}-\textbf{R}\right|^{2}}{4D_{\varepsilon}t}\right),\label{dp}\end{equation}
 where $D_{\varepsilon}$ is the (energy dependent) diffusion coefficient
(see Sec. 3). When $\epsilon$ approaches the mobility edge $E_{c}$,
the disorder gets stronger and (\ref{dp}) is not applicable any longer.
For that case one can use the scaling theory for the Fourier transformed
kernel, $P_{\varepsilon}(\textbf{r},\textbf{R},\Omega)$, which is
related to the dynamic diffusion coefficient $D_{\varepsilon}(\Omega)$
\cite{Lee}.

We start with the ordinary diffusion, (\ref{dp}), and restrict ourselves
to the long time limit, when (\ref{App3}) is valid. If $D$ were
constant, i.e. energy independent, (\ref{App3}) would reduce to a
Gaussian function (the momentum distribution is normalized to the
total number of particles $\int\frac{d\textbf{k}}{(2\pi)^{d}}|\tilde{\phi}(\textbf{k})|^{2}=N$).
Due to the energy dependence of $D$, the shape $\overline{n}(\textbf{r},t)$
deviates from Gaussian and it depends on the initial momentum distribution
$|\tilde{\phi}(\textbf{k})|^{2}$. The characteristic width of this
distribution is $k_{\mu}=\sqrt{2m\mu/\hbar^{2}}=1/\xi_{h}$ and it
is due to the rapidly oscillating part which the wave function acquires
after the first stage of a ballistic expansion. To a good approximation,
in any space dimension, the $k$-dependence of $|\tilde{\phi}(\textbf{k})|^{2}$
is determined by the factor $(1-\frac{k^{2}}{2k_{\mu}^{2}})$, \cite{Kag,Cas,Mini}.
For instance, in $2d$ \begin{equation}
|\tilde{\phi}(\textbf{k})|^{2}=\frac{4\pi N}{k_{\mu}^{2}}\left(1-\frac{k^{2}}{2k_{\mu}^{2}}\right)\Theta(k_{\mu}\sqrt{2}-k).\label{mom}\end{equation}

Let us consider a short-range correlated potential, i.e. $\eta<<1$,
and assume that $k_{\mu}R_{0}<<1$. It must be also assumed that the
chemical potential $\mu$ of the condensate (prior to the release
from the trap) is much above the mobility edge $E_{c}$, so that the
great majority of the $k$-components experience weak disorder and
propagate by diffusion. 
In $2d$ the stated conditions are the same as given in the inequalities
in the first line of (\ref{D2}). The mean free time is then independent
of $k$ and the diffusion coefficient is $D_{k}=\hbar^{2}k^{2}\tau/2m^{2}=D_{\mu}(k/k_{\mu})^{2}$,
where $D_{\mu}$ is the diffusion coefficient for $k=k_{\mu}$.

The upper limit in the integral (\ref{App3}) is $k_{\mu}\sqrt{2}$.
Since for sufficiently small $k$ diffusion breaks down, a lower
cutoff, $k_{c}$, should be introduced in the integral. The cutoff
is estimated from the condition $kl=1$ which, using (\ref{mfp2}),
gives $k_{c}=(\eta/R_{0})=(m/\hbar\tau)^{1/2}$. This cutoff is a
rough benchmark between the components which diffuse away and
those which stay localized. Writing the kernel (\ref{dp}) in terms
of $k$ and calculating the integral in (\ref{App3}), with the
expression (\ref{mom}) for $|\tilde{\phi}(\textbf{k})|^{2}$ and
with a lower limit $k_{c}$, yields \begin{equation}
\overline{n}(\textbf{r},t)=\frac{N}{4\pi D_{\mu}t}\left\{
(1+\alpha)\left[E_{1}(\alpha)-E_{1}
(\frac{2\alpha}{\delta^{2}})\right]+\frac{1}{2}\delta^{2}\exp(-\frac{2\alpha}{\delta^{2}})-e^{-\alpha}\right\}
,\label{shape1}\end{equation}
 where $E_{1}(x)$ is the exponential integral, $\alpha\equiv r^{2}/8D_{\mu}t$
and $\delta=(k_{c}/k_{\mu})=\sqrt{\hbar/2\mu\tau}$ is the dimensionless
cutoff. The cutoff is important only for small $r$, to avoid the
divergence at $r=0$. For $r^{2}>>\hbar t/m$ the parameter $2\alpha/\delta^{2}$
is large and the two $\delta$-dependent terms in (\ref{shape1})
can be neglected.

In Fig.2 we plot the normalized particle density, $\tilde{n}(\textbf{r})=4\pi D_{\mu}t\overline{n}(\textbf{r},t)/N$,
for some fixed time $t$, as a function of the normalized distance
$\tilde{r}=\frac{r}{\sqrt{8D_{\mu}t}}$. %
\begin{figure}[H]
\centering{}\includegraphics[scale=0.4]{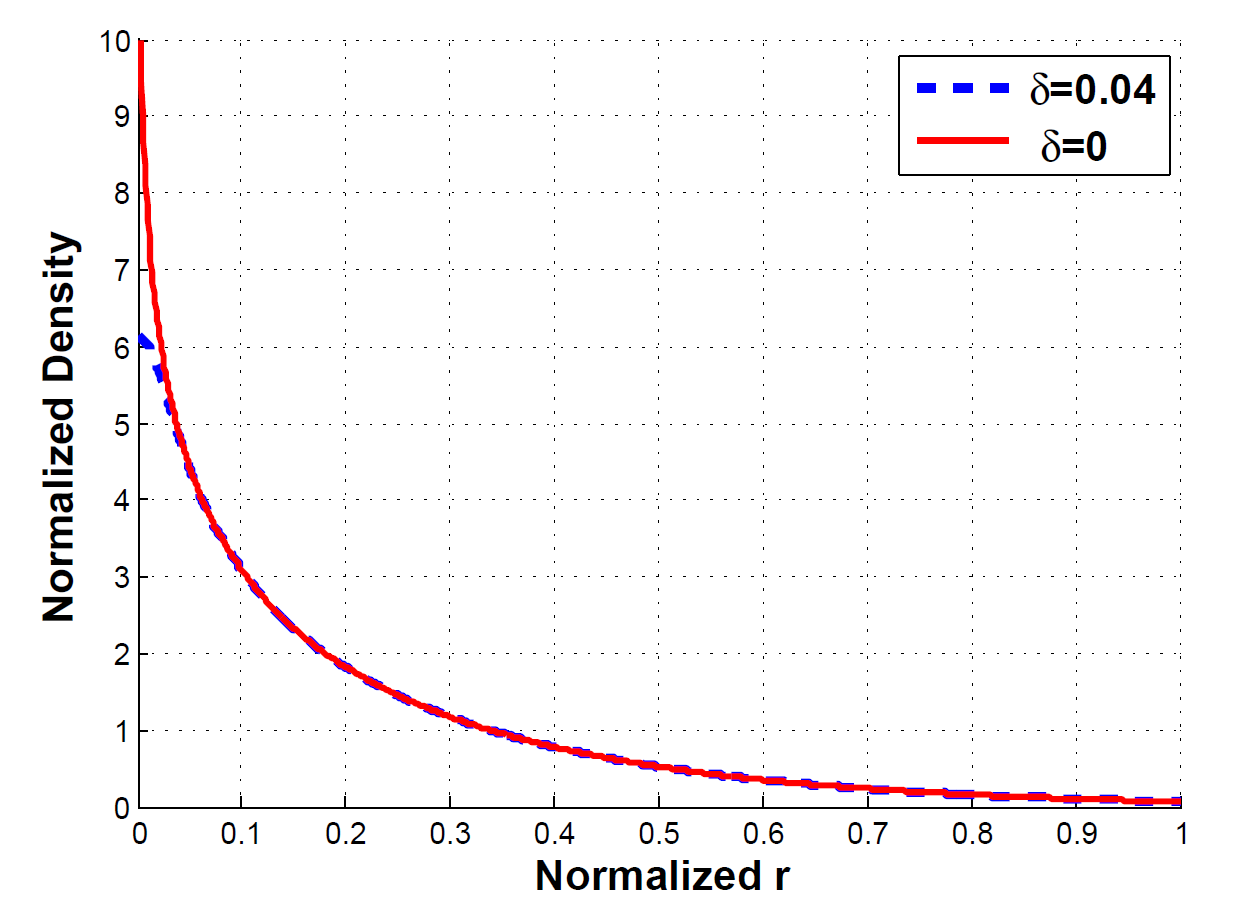}\caption{Density as a
function of position (both quantities are normalized as explained
in the text). The cutoff $\delta$ is zero for the solid curve and
it is $0.04$ for the dashed curve. }

\end{figure}
The dashed curve corresponds to a cutoff $\delta=0.04$ while for
the solid curve $\delta=0$. The curves exhibit an exponential tail
(with a power-law prefactor) for large $\tilde{r}$. This tail
comes from the large-$k$ part of $|\tilde{\phi}(\textbf{k})|^{2}$,
i.e. from particles which, in a given time, manage to diffuse far
away. It can be obtained analytically, using the large-$z$
asymptotics $E_{1}(z)\approx z^{-1}\exp(-z)$. The sharp increase
in $\tilde{n}(\textbf{r})$ for small $\tilde{r}$ is due to the
small-$k$ components whose diffusion coefficient decreases as
$k^{2}$. Since for small $z$, $E_{1}(z)\approx -\ln z$, the solid
curve has a logarithmic divergence at the origin. The divergence
is rounded off by the finite cutoff $\delta$.

It is sometimes convenient to replace the sharp cutoff in
(\ref{mom}) by a smoother, Gaussian cutoff, thus, taking
$|\tilde{\phi}(\textbf{k})|^{2}=(4\pi
N/k_{\mu}^{2})\exp(-k^{2}/k_{\mu}^{2})$, with the result
\cite{BS1}: \begin{equation}
\overline{n}(\textbf{r},t)=\frac{N}{2\pi
D_{\mu}t}K_{0}\left(\frac{r}{\sqrt{D_{\mu}t}}\right),\label{shape}\end{equation}
 where $K_{0}$ is the zeroth order modified Hankel function, and
no lower cutoff $\delta$ was introduced. This result is very similar
to (\ref{mom}), with $\delta=0$. Indeed, 
 the function $K_{0}(z)$ has an exponential tail (with a power-law
prefactor) for large $z$, and it diverges logarithmically for small
$z$. 

It is interesting that $D_{\mu}$ is the only system-dependent
quantity (except for the total number of atoms $N$) which appears
in (\ref{shape1}) or (\ref{shape}). Using the natural quantum unit
of diffusion, $\hbar/m$, one can define a dimensionless parameter
$mD_{\mu}/\hbar\sim\mu\tau^{*}/\hbar$, where $\tau^{*}$ is the
transport mean free time for a particle with energy $\mu$ (this
relation is not restricted to the above $2d$ example). This
parameter carries information about the initial state of the
condensate ($\mu$ depends on the nonlinearity and on other
factors), as well as about the subsequent dynamics (via
$\tau^{*}$). It is analogous to the parameter $E_{F}\tau/\hbar$
which determines the transport properties of a disordered metal
($E_{F}$ is the Fermi energy).

Diffusion of cold $^{87}Rb$ atoms, in a $2d$-geometry, was
observed experimentally in \cite{MR}. Motion in the vertical ($z$)
direction was confined and transport in the $x,y$-plain was
studied. The temperature was somewhat above the condensation
temperature, so that instead of the inverted parabola (\ref{mom})
one should use the appropriate thermal distribution \cite{MR}.
Since the shape of the diffusing cloud is not very sensitive to
the precise initial distribution, provided it still contains fast
and slow particles, one can expect results similar to those in
Fig.2. Indeed, profiles of the (integrated along $x$ or $y$)
density, shown in Fig.3, have clear resemblance with Fig.2. Note
the difference between curves in (b) and (c) of Fig.3, for the
same propagation time. It is due to the anisotropy in the
potential and hence in the diffusion coefficient. The experimental
data in \cite{MR} were fit numerically to a model with a
anisotropic, energy-dependent diffusion coefficient.
\begin{figure}[H]
\centering{}\includegraphics[scale=0.45]{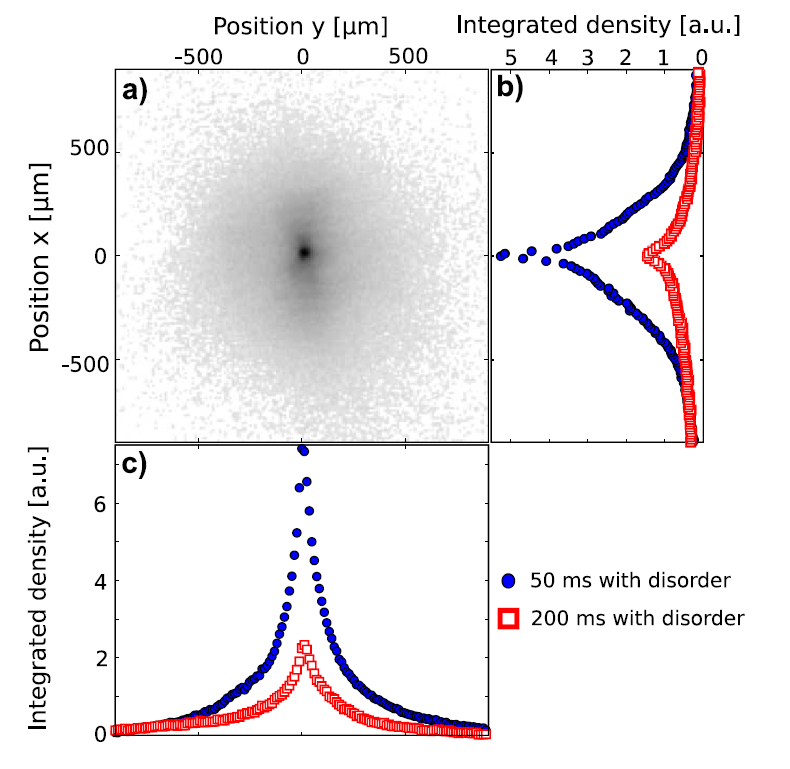}\caption{ Atomic
column density after planar expansion of an ultracold gas in an
anisotropic speckle potential. (a) Image after 50 ms of expansion.
(b), (c) Integrated density along the two major axes. The plain
dots (open squares) correspond to 50 ms (200 ms) of expansion.
(Reprinted, with permission, from Phys. Rev. Lett. \textbf{104},
220602 (2010)).}

\end{figure}

The characterization of a diffusing atomic cloud is not limited to
the average particle density $\overline{n}(\textbf{r},t)$. Indeed,
in a \textit{given realization} of randomness the density pattern
$n(\textbf{r},t)$ exhibits spatial fluctuations, which are smoothed
out only in the process of averaging over many realizations. The fluctuations
are due to multiple scattering of the matter wave on the random potential.
The phenomenon is similar to optical speckles for the electromagnetic
waves (see Sec. 2) and can be termed \textquotedbl{}matter wave speckles\textquotedbl{}.
The density fluctuations can be described by the (equal time) two
point density correlation function, $\overline{n(\boldsymbol{r},t)n(\boldsymbol{r}',t)}$,
as well as by higher order correlation functions. The function $\overline{n(\boldsymbol{r},t)n(\boldsymbol{r}',t)}$
reveals both short and long-range correlations, which ( for a white
noise potential) have been studied, respectively, in \cite{Hens}
and \cite{Cher2}.

Finally, let us recall that our discussion has been limited to the
\textquotedbl{}two-stage scenario\textquotedbl{}: first, a free ballistic
expansion and only later, when the nonlinearity had decreased sufficiently
, the disorder is switched on. An interesting and conceptually important
question is to what extent the diffusion is modified in the simultaneous
presence of disorder and nonlinearity, i.e. when a wave packet evolves
according to (\ref{G3}). This problem has been addressed in \cite{Cher,Schwiete,Cher1}.
It was shown \cite{Cher} that in the first order in perturbation
theory the nonlinearity causes a slight renormalization of the diffusion
coefficient which controls the condensate spreading. In the works
\cite{Schwiete,Cher1}, which went far beyond the simple perturbation
theory, a generalized diffusion equation for the spectrally resolved
(average) density $\overline{n}_{\epsilon}(\textbf{r},t)$ was derived.
The \textquotedbl{}diffusion coefficient\textquotedbl{} in that equation
depends on $\textbf{r}$ and $t$, through the total particle density
$\overline{n}(\textbf{r},t)$, which has to be determined self-consistently.

\subsection{Localization of a BEC}

The general expressions (\ref{n1 av}), (\ref{n2 av}), derived in
the previous subsection, apply also for strong disorder (i.e. when
$\epsilon$ is in the vicinity, or below the mobility edge
$E_{c}$), provided one uses the appropriate expressions for the
propagation kernel and the spectral function. It is convenient to
define the Fourier transform, $P_{\epsilon}(\textbf{r},\Omega)$,
of the kernel $P_{\epsilon}(\textbf{r},0,t)\equiv
P_{\epsilon}(\textbf{r},t)$. For ordinary diffusion, (\ref{dp}),
$P_{\epsilon}(\textbf{r},\Omega)=(4\pi
D_{\epsilon}r)^{-1}\exp(-r\sqrt{-i\Omega/D_{\epsilon}})$. More
generally, \begin{equation}
P_{\epsilon}(\textbf{r},\Omega)=\frac{1}{4\pi
D_{\epsilon}(\Omega)r}\exp\left[-r\sqrt{\frac{-i\Omega}{D_{\epsilon}(\Omega)}}\;\right],\label{P_Omega}\end{equation}
 where $D_{\epsilon}(\Omega)$ is the dynamic, frequency dependent
diffusion coefficient which carries information about transport beyond
ordinary diffusion. This quantity has been extensively studied in
the context of disordered electronic systems, where it is proportional
to the $ac$ conductivity $\sigma_{\epsilon}(\Omega)$, with $\epsilon$
being the Fermi energy \cite{Lee,Chalk}. The $\Omega\rightarrow0$
limit corresponds to the $dc$ value. In the localized regime, $\epsilon<E_{c}$,
this value is zero, $D_{\epsilon}(0)\equiv D_{\epsilon}=0$. The dynamic
diffusion coefficient, however, is not zero and, for small $\Omega$,
it is given by \begin{equation}
D_{\epsilon}(\Omega)\sim-i\Omega\xi^{2}(\epsilon),\;\;\;\;\;\;\;\;\;\;\;\;(\epsilon<E_{c})\label{D1}\end{equation}
 where $\xi$ is the localization length. Close to the transition
$\xi(\epsilon)\propto(E_{c}-\epsilon)^{-\nu}$, where the exponent
$\nu\approx1.57$ \cite{Slevin}. The expression (\ref{D1}) is easily
understood in terms of $\sigma_{\epsilon}(\Omega)$: although there
is no $dc$ transport (the eigenstates at energy $\epsilon$ are localized),
an external field at frequency $\Omega$ polarizes the medium, causing
an oscillating polarization current.

Note that (\ref{D1}) is valid only for $\Omega<\Omega_{c}(\xi)$,
where the crossover frequency $\Omega_{c}(\xi)$, estimated below,
decreases to zero when $\xi\rightarrow\infty$. At the transition
point, $\epsilon=E_{c}$, (\ref{D1}) has no region of applicability
and it is replaced by $D_{c}=(-i\gamma\Omega)^{1/3}$, where $\gamma$
is a model-dependent coefficient. This behavior follows from the scaling
theory of localization \cite{Weg,Ab,Lee} and it implies that at the
mobility edge a wave packet spreads according to $\overline{r^{2}}\propto t^{2/3}$
(anomalous diffusion), rather than $\overline{r^{2}}\propto t$ (
ordinary diffusion). Comparing $D_{c}$ with $D_{\epsilon}(\Omega)$
in (\ref{D1}) yields $\Omega_{c}(\xi)\sim\sqrt{\gamma}/\xi^{3}$,
or a characteristic time $t_{c}=(1/\Omega_{c})\propto\xi^{3}$. This
is the crossover time between anomalous diffusion and localization:
A particle (a wave packet) with $\epsilon$ somewhat below $E_{c}$
will propagate by anomalous diffusion up to time of order $t_{c}$
and only then (i.e. after it had spread over a region of size $\xi$)
will it \textquotedbl{}realize\textquotedbl{} that it is in fact localized.

A similar crossover occurs at the other side of the transition, $\epsilon>E_{c}$,
where the eigenstates are extended. Close to the transition there
exists a macroscopic length $\xi(\epsilon)\propto(\epsilon-E_{c})^{-\nu}$,
with the meaning of being a crossover length between anomalous and
ordinary diffusion. A particle with an energy slightly above $E_{c}$
first experiences anomalous diffusion, up to a scale $\sim\xi$, and
only at a larger scale exhibits ordinary diffusion, with a small $dc$
diffusion coefficient $D_{\epsilon}=1/\hbar\nu_{c}\xi\propto(\epsilon-E_{c})^{\nu}$
(here $\nu_{c}$ is the density of states at the energy $E_{c}$).
Thus, slightly above the transition one can view $D_{\epsilon}(\Omega)$
as a sum of two contributions: the $dc$ part and the anomalous part,
proportional to $(-i\Omega)^{1/3}$. On the other hand, well above
$E_{c}$, where the disorder is weak, the diffusion coefficient (up
to small $\Omega$-dependent corrections, responsible for the weak
localization effects) is given by the expressions (\ref{D}) or (\ref{DD}),
depending on the value of the parameter $\eta$. The existence of
a macroscopic length $\xi(\epsilon)\propto(|\epsilon-E_{c}|^{-\nu}$
on \textit{both} sides of the transition is a consequence of scaling
and it is quite natural: only by observing the system at a scale larger
than $\xi$ can one distinguish between localized and extended states.
Using the expressions for $D_{\epsilon}(\Omega)$ in various limits,
and possibly extrapolating between them, one can reconstruct, via
(\ref{P_Omega}), the quantum diffusion kernel $P_{\epsilon}(\textbf{r},t)$.
A useful interpolation scheme is the self-consistent theory \cite{Voll},
employed in the numerical part of \cite{Skip}. Its drawback, though,
is that it yields a wrong value for the exponent $\nu$ (1 instead
of approximately 1.57, in 3d).

In order to use (\ref{App2}) one has to know the spectral function,
in addition to the quantum diffusion kernel. The average Green's function,
in the momentum representation, can be written as \cite{Akker} (the
Dyson equation): \begin{equation}
\overline{G}(\textbf{k},\epsilon)=\frac{1}{\epsilon-\epsilon_{\textbf{k}}-\sum(\textbf{k},\epsilon)}\;\;,\label{Born1}\end{equation}
 where $\sum(\textbf{k},\epsilon)$ is the self-energy. The simplest
approximation for the self- energy is the first order Born approximation:
$\sum(\textbf{k},\epsilon)=V_{0}^{2}\int(2\pi)^{-3}d^{3}k'\tilde{\Gamma}(\textbf{k}-\textbf{k}')G_{0}(\textbf{k}',\epsilon)$,
where $\tilde{\Gamma}(\textbf{q})$ is the Fourier transform of the
potential correlation function $\Gamma(\textbf{R}/R_{0})$ and $G_{0}(\textbf{k},\epsilon)$
is the unperturbed (i.e. in the absence of disorder) Green's function.
A more elaborate approximation is the self-consistent Born approximation,
which amounts to replacing $G_{0}(\textbf{k},\epsilon)$ by the full
Green's function, (\ref{Born1}), thus, obtaining a closed equation
for the self- energy: \begin{equation}
\sum(\textbf{k},\epsilon)=V_{0}^{2}\int\frac{d^{3}k'}{(2\pi)^{3}}\frac{\tilde{\Gamma}(\textbf{k}-\textbf{k}')}{\epsilon-\epsilon_{\textbf{k}'}-\sum(\textbf{k}',\epsilon)}.\label{Born2}\end{equation}
 The solution of this equation is, obviously, an even function of
$V_{0}$. Therefore the self-consistent Born approximation is hardly
appropriate for the speckle potential, where the odd powers of $V_{0}$
are known to be generally important \cite{Kuhn}. 
An analytic solution of (\ref{Born2}) can be obtained \cite{Skip}
in the limit of a short range potential, when $V_{0}^{2}\Gamma(\textbf{R}/R_{0})$
can be replaced with $u\delta(\textbf{R})$, so that $\tilde{\Gamma}(\textbf{k}-\textbf{k}')=u$.
The constant $u$ is proportional to $V_{0}^{2}R_{0}^{3}$ and the
formal limit $V_{0}\rightarrow0,R_{0}\rightarrow\infty,V_{0}^{2}R_{0}^{3}=const$
is implied. For instance, for $\Gamma=exp(-R^{2}/R_{0}^{2})$ (see
Sec. 3.1), $u=\pi\sqrt{\pi}V_{0}^{2}R_{0}^{3}$. The solution of (\ref{Born2}),
with $\tilde{\Gamma}(\textbf{k}-\textbf{k}')=u$, yields a $\textbf{k}$-independent
self-energy \begin{equation}
\sum(\epsilon)=\epsilon^{*}+\frac{\hbar^{2}}{8ml^{2}}-\frac{i\hbar}{2\tau_{\epsilon}}\;\;.\label{Born}\end{equation}
 The value of $\epsilon^{*}$ depends on the ultraviolet cutoff, needed
to regularize the integral in (\ref{Born2}). For small but finite
correlation radius $R_{0}$ the natural cutoff is $1/R_{0}$, which,
in the leading order in the small parameter $\eta$, yields $\epsilon^{*}\sim-\eta V_{0}$.
The mean free path (in $3d$) is $l=\pi\hbar^{4}/mu^{2}$ and it
does not depend on $\epsilon$. This expression for $l$ is entirely
consistent with the result $l=4R_{0}/\eta^{2}\sqrt{\pi}$ of Sec.
3.1 (see the line below (\ref{l''})), if one substitutes the value
$u=\pi\sqrt{\pi}V_{0}^{2}R_{0}^{3}$. The mean free time
$\tau_{\epsilon}$ is related to $l$ in the usual way,
$\tau_{\epsilon}=ml/\hbar k$, with $\hbar
k=\sqrt{2m(\epsilon-\epsilon^{*})}$. From the condition $kl=1$ one
obtains for the mobility edge
$E_{c}=\epsilon^{*}+(\hbar^{2}/2ml^{2})$. For $\eta$ small,
$E_{c}$ is negative (see the corresponding remark in Sec. 3.1). In
what follows we set $\epsilon^{*}=0$ (the edge of
the spectrum) 
 and count all the energies from this point. The mobility edge becomes
$E_{c}=\hbar^{2}/2ml^{2}$.

Knowledge of the self-energy enables us to write down the expression
(\ref{Born1}) for $\overline{G}(\textbf{k},\epsilon)$, and hence
for the spectral function \begin{equation}
-\frac{1}{\pi}Im\overline{G}(\textbf{k},\epsilon)\equiv A(\textbf{k},\epsilon)=\frac{1}{\pi E_{c}}\frac{\sqrt{x}}{(x-\frac{1}{4}-\frac{\hbar^{2}k^{2}}{2mE_{c}})^{2}+x}\;\;,\;\;\;\;\;\;\;\;\;\; x=\frac{\epsilon}{E_{c}},\label{spectral}\end{equation}
 where, instead of $\epsilon$, the dimensionless variable $x$ has
been introduced.



Let us fix some distant point $\textbf{r}$ and observe the atoms
arriving at this point in the course of time, according to (\ref{App2}).
If the mobility edge $E_{c}$ is well below the chemical potential
$\mu$ (recall that information about $\mu$ in (\ref{App2}) is carried
by the momentum distribution $\tilde{\phi}(\textbf{k})$), then $\overline{n}(\textbf{r},t)$,
as a function of time, will exhibit several regimes. First, the fastest
atoms will arrive, by ordinary diffusion (Sec. 5.2) and the density
will reach a maximum at a time $t_{arrival}\simeq(r^{2}/6D_{\mu})$.
Later, slower particles, with $\epsilon$ closer to $E_{c}$, 
will show up. Still later, the anomalously diffusing particles will
make their appearance and, finally, for times much larger than the
characteristic time $t_{loc}\simeq(\hbar r^{3}/E_{c}l^{3})$ only
the localized part of the condensate will remain. All this has been
studied in some detail in \cite{Skip}. We discuss only the $t\rightarrow\infty$
limit, when only the localized part of the condensate is left, at
any finite $\textbf{r}$. In this limit $D_{\epsilon}(\Omega)$ is
given by (\ref{D1}) which, upon substitution into (\ref{P_Omega})
and Fourier transforming with respect to $\Omega$, yields the kernel
\begin{equation}
P_{\epsilon}(\textbf{r},\textbf{R},t\rightarrow\infty)=\frac{1}{4\pi|\textbf{r}-\textbf{R}|\xi^{2}}\exp\left(-\frac{|\textbf{r}-\textbf{R}|}{\xi}\right).\label{kernel
loc}\end{equation}
 The density $\overline{n}(\textbf{r},t\rightarrow\infty)\equiv\overline{n}_{loc}(\textbf{r})$
is then obtained from (\ref{n2 av}) by cutting the upper limit of
the integral over $\epsilon$ at $\epsilon=E_{c}$. Since close to
$E_{c}$ the localization length $\xi(\epsilon)$ is large, the localized
part of the condensate exhibits long tails. For $r$ well away from
the initial location of the condensate, and making use of (\ref{spectral}),
one obtains \cite{Skip}: \begin{equation}
\overline{n}_{loc}(\textbf{r})\sim f(E_{c}/\mu)\frac{N}{r^{3}}\left(\frac{l}{r}\right)^{1/\nu},\label{nloc}\end{equation}
 where $f(z)\sim z^{3/2}$ for $z<<1$ and $f(z)$ approaches a constant
for $z>>1$. This localized tail contains information about the localization
length exponent $\nu$.

An important quantity is the (average) number of localized atoms,
$\overline{N}_{loc}$, or the fraction of the condensate, $N_{loc}/N$,
which remains localized after the mobile part had diffused away. The
expression for $\overline{N}_{loc}$ is derived from (\ref{n2 av}),
by integrating it over $\textbf{r}$ and setting the upper limit in
the integral over $\epsilon$ at $E_{c}$. Since integration of the
kernel $P_{\epsilon}(\textbf{r},t)$ over $\textbf{r}$ gives unity,
one obtains \begin{equation}
\overline{N}_{loc}=\int\frac{d\textbf{k}}{(2\pi)^{d}}|\tilde{\phi}(\textbf{k}|^{2}\int_{-\infty}^{E_{c}}d\epsilon A(\textbf{k},\epsilon).\label{Nloc}\end{equation}
 In view of the importance of this equation we give an independent,
and rather general, derivation. Denoting by $\epsilon_{\alpha},|\alpha>$
the energy eigenvalues and eigenvectors for a given realization of
disorder, one can write $\Psi(\textbf{r},t)\equiv<\textbf{r}|\Psi>=\sum_{\alpha}<\textbf{r}|\alpha><\alpha|\Phi>exp(-i\epsilon_{\alpha}t)$,
where $\Phi$ is the initial wave function. In the $t\rightarrow\infty$
limit, for any fixed value of $\textbf{r}$, only localized eigenstates
contribute to this expression. Taking its square modulus, integrating
over \textbf{r} and using the orthonormality of the $\alpha$-basis,
yields a simple, intuitively obvious expression for the number of
localized atoms $N_{loc}=\sum'|<\alpha|\Phi>|^{2}$, where the prime
indicates that the summation is over localized states only. This expression
can be written as \begin{equation}
N_{loc}=\int_{-\infty}^{E_{c}}d\epsilon\sum_{\alpha}\delta(\epsilon-\epsilon_{\alpha})<\Phi|\alpha><\alpha|\Phi>=-\frac{1}{\pi}\int_{-\infty}^{E_{c}}d\epsilon<\Phi|Im\hat{G}(\epsilon)|\Phi>,\label{Nloc1}\end{equation}
 where $\hat{G}(\epsilon)=(\epsilon-\hat{H})^{-1}$ is the resolvent
operator (for the given disorder realization). Writing (\ref{Nloc1})
in the momentum representation and averaging over the disorder results
in Eq. (\ref{Nloc}). $|\tilde{\phi}(\textbf{k})|^{2}$ in this equation
is a function of $k/k_{\mu}$. It assumes a constant value $\sim N/(k_{\mu})^{3}$
for $k<<k_{\mu}$, with a cutoff at $k\sim k_{\mu}$. Furthermore,
$A(\textbf{k},\epsilon)$, as a function of $k$, decays on a scale
$k_{c}\sim\sqrt{2mE_{c}}/\hbar$ (see (\ref{spectral})). It follows
then from (\ref{Nloc}) that $N_{loc}$ is a function of the parameter
$k_{\mu}/k_{c}$ (or $\mu/E_{c}$). We will not compute the detailed
shape of this function but rather analyze the two limits:

(i) $k_{\mu}>>k_{c}$. In this case the effective region of integration
is $k\sim k_{c}$, so that the function $|\tilde{\phi}(\textbf{k})|^{2}$
can be replaced by its value at $k=0$, and the remaining integral
over \textbf{k} gives the density of states $\nu(\epsilon)$, i.e.
\begin{equation}
N_{loc}=|\tilde{\phi}(k=0)|^{2}\int_{-\infty}^{E_{c}}d\epsilon\nu(\epsilon)\sim N(k_{c}/k_{\mu})^{3}.\label{Nloc2}\end{equation}
 The fraction of localized atoms $N_{loc}/N<<1$.

(ii) $k_{\mu}<<k_{c}$. Now the spectral function is \textquotedbl{}flat\textquotedbl{}
in the effective integration region, $k<k_{\mu}$, and $\textbf{k}$
in $A(\textbf{k},\epsilon)$ can be set to $0$. The remaining integral
over \textbf{k} in (\ref{Nloc}) is equal to $N$, by normalization,
and using (\ref{spectral}), one obtains: \begin{equation}
\frac{N_{loc}}{N}\equiv f_{loc}=\frac{1}{\pi}\int_{0}^{1}dx\frac{\sqrt{x}}{(x+\frac{1}{4})^{2}}\approx0.45.\label{Nloc3}\end{equation}
 This result is surprising since the natural expectation is that,
in the limit of strong disorder, the fraction of localized atoms should
approach unity. A short discussion of this result is given in \cite{Skip}.
The issue deserves further investigation, especially since (\ref{spectral})
becomes doubtful in the strong disorder limit.

Recently an experimental observation of Anderson localization of a
BEC in $3d$ has been reported \cite{Jen}. A dilute $^{87}Rb$
condensate was released from an optical trap and allowed to freely
expand for a time $t_{0}=50ms$. After that time the random speckle
potential was suddenly switched on and the dynamics of the
condensate in that potential has been studied, i.e. images of the
atomic cloud at various times have been taken. In Fig. 4 two
sequences of images are shown (the time is measured from the
instance when the disorder was switched on). The upper sequence
corresponds to weak disorder, when a large fraction of the
condensate evolves by diffusion. In the lower sequence
disorder is strong and a significant fraction of atoms remains localized.%
\begin{figure}[H]
\centering{}\includegraphics[scale=0.5]{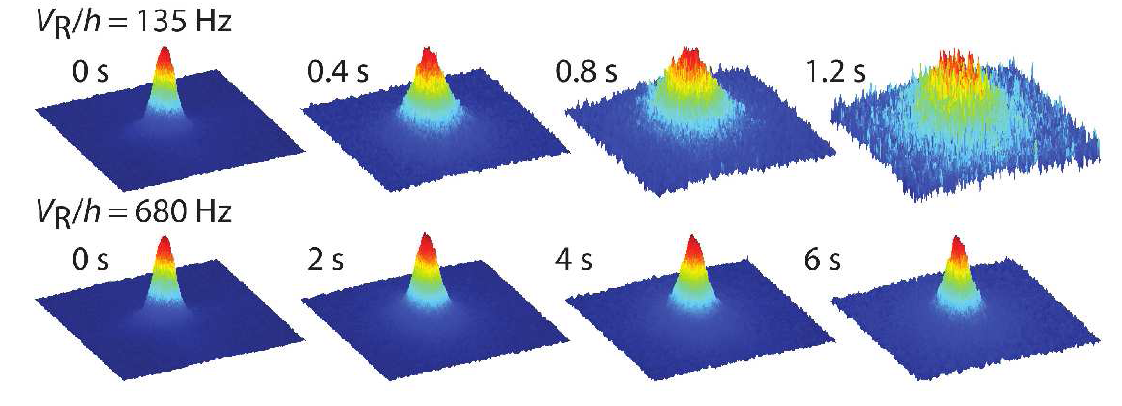}\caption{$3d$
expansion of the atomic cloud for the two values of the disorder
amplitude. For small disorder ($V_0/h = 135 Hz$) one observes an
essentially diffusive time evolution. For larger disorder ($V_0/h
= 680 Hz$) a significant fraction of atoms remains localized.
(Reprinted from the arXiv: 1108.0137.)}

\end{figure}
 The theory presented in \cite{Jen} starts with the following expression
for the average density of atoms: \begin{equation}
\overline{n}(\textbf{r},t)=\int d\textbf{R}\int d\epsilon P_{\epsilon}(\textbf{r},\textbf{R},t){\textsl{D}}(\textbf{R},\epsilon)\;\;,\label{jer}\end{equation}
 where the function ${\textsl{D}}(\textbf{R},\epsilon)$ is interpreted
as the semiclassical density in the position-energy space. Note that
with ${\textsl{D}}(\textbf{R},\epsilon)\equiv\int d\textbf{k}W(\textbf{k},\textbf{R})A(\textbf{k},\epsilon)$
$\;$ Eq. (\ref{jer}) is compatible with (\ref{n2 av}). For strong
disorder, $E_{c}>>\mu$, the spectral function is a broad, slowly
decaying function of $k$, and it can be replaced by its $k$-zero
value, $A(k=0,\epsilon)\equiv f(\epsilon)$. The remaining integral
over $\textbf{k}$ gives the initial (i.e. at time $t_{0}$ ) density
$n_{0}(\textbf{r})$, so that ${\textsl{D}}(\textbf{R},\epsilon)=n_{0}(\textbf{r})f(\epsilon)$.
A further approximation made in \cite{Jen} was to replace $P_{\epsilon}(\textbf{r},\textbf{R},t)$
by $\delta(\textbf{r}-\textbf{R})$, thus, obtaining for the localized
part of the condensate \begin{equation}
\overline{n}_{loc}(\textbf{r})=n_{0}(\textbf{r})\int_{-\infty}^{E_{c}}d\epsilon f(\epsilon)\equiv n_{0}(\textbf{r})f_{loc}\;\;,\label{jer1}\end{equation}
 which is just the initial density profile, rescaled by the factor
$f_{loc}$. The function $f(E)$ was calculated in \cite{Jen} numerically
by diagonalizing the Hamiltonian, for the realistic anisotropic speckle
potential, and averaging over several realizations. A value $f_{loc}\approx0.22$
was obtained, in the strong disorder limit (compare to $0.45$ in
(\ref{Nloc3}), for the white noise disorder), and after some heuristic
energy shift a fair agreement with the experimental results was achieved.
Quantitative theoretical results are quite sensitive to various approximations
made and, in particular, we note that replacing $P_{\epsilon}(\textbf{r},\textbf{R},t)$
by $\delta(\textbf{r}-\textbf{R})$, which might be reasonable for
strongly localized states, will break down close to $E_{c}$ where
the localization length becomes large. Therefore (\ref{jer1}) for
the localized density profile cannot be the full story: $\overline{n}(\textbf{r},t\rightarrow\infty)$
should have power law decaying tails (see (\ref{nloc})), associated
with energies close to $E_{c}$.

Let us also mention a different type of localization, which takes
place in the momentum space, rather than in real space. When a
quantum system (chaotic in the classical limit) is
\textquotedbl{}kicked\textquotedbl{} by a sequence of short
pulses, the expectation value of the momentum square,
$<p^{2}(t)>$, evolves in time. In the well studied kicked rotor
model $<p^{2}(t)>$ saturates in the long time limit \cite{Fish}- a
phenomenon known as dynamical localization, which was observed
experimentally in a system of cold atoms \cite{Raiz}.  A $3d$
extension of the model \cite{Cass} exhibits a richer behavior,
analogous to the Anderson transition: Depending on the kicking
strength, $K$, $<p^{2}(t)>$ can either linearly grow with time
(diffusion), saturate in the long time limit (dynamical
localization) or display the above mentioned anomalous diffusion,
$<p^{2}(t)>\;\propto t^{2/3}$ (critical behavior at the Anderson
transition point). The latter occurs at a critical value, $K_{c}$,
of the kicking strength. Such \textquotedbl{}dynamical Anderson
transition\textquotedbl{} has been experimentally observed in
\cite{Del1,Del2,Del3}. A system of cold cesium atoms was acted
upon by a pulsed laser wave. The authors have identified the
transition point $K_{c}$, measured the localization parameter,
$(K_{c}-K)^{-\nu}$, close to the transition, and extracted the
value of the exponent $\nu$, in very good agreement with the
accepted numerical value.

\subsection{Cold Fermi gas}

Cold Fermi gases are also of interest, as reviewed in \cite{Giorgini,YCastin}.
Although interaction between the fermions are generally important,
in some cases they have only a minor effect on the dynamics of the
atomic cloud. This is particularly true for a polarized Fermi gas
when the Pauli principle suppresses the main mechanism (the s-scattering)
for the interaction. In this subsection we briefly review some phenomena
pertaining to the expansion of a cold Fermi gas released from a harmonic
trap.

In the trap, the fermions occupy the single-particle states,
$\phi_{n}(\textbf{R})$, of the harmonic oscillator potential
$m\omega^{2}R^{2}/2$. The derivation of the expression for the
average particle density $\overline{n}(\textbf{r},t)$ proceeds as
in Sec. 5.2. The only difference is that, instead of the
condensate wave function $\Phi(\textbf{R})$, we now have $N$
incoherent single-particle contributions. At zero temperature,
when the fermions occupy states up to the Fermi energy $E_{F}$,
the corresponding Wigner function is \begin{equation}
W_{F}(\textbf{k},\textbf{R})=\frac{1}{(2\pi)^{d}}\Theta\left(E_{F}-\frac{\hbar^{2}k^{2}}{2m}-\frac{1}{2}m\omega^{2}R^{2}\right),\label{Fermi1}\end{equation}
 and the dynamics of the gas, upon its release from the trap, is described
by (\ref{n2 av}), with $W_{F}(\textbf{k},\textbf{R})$ instead of
$W(\textbf{k},\textbf{R})$. In particular, for weak disorder, when
the spectral function can be approximated by $\delta(\epsilon-\epsilon_{\textbf{k}})$,
one obtains \cite{Beilin} \begin{equation}
\overline{n}(\textbf{r},t)=\int d\textbf{R}\int\frac{d\textbf{k}}{(2\pi)^{d}}P_{k}(\textbf{r},\textbf{R},t)\Theta\left(E_{F}-\frac{\hbar^{2}k^{2}}{2m}-\frac{1}{2}m\omega^{2}R^{2}\right),\label{AppF1}\end{equation}
 which is the precise analog of (\ref{App1}). Diffusion of a Fermi
cloud in a $2d$ speckle potential was analyzed in \cite{Beilin}.
The authors used (\ref{AppF1}) with the diffusion propagator (\ref{dp})
and with $D_{\epsilon}$ appropriate for the $2d$ speckle \cite{Mini}.
The density profile $\overline{n}(\textbf{r},t)$ turns out to be
very similar to that for a diffusing BEC (compare (\ref{shape1})
of Sec. 5.2 with Eq.(26) in \cite{Beilin}). This coincidence is not
surprising because the Wigner function (\ref{Fermi1}) for fermions
has much in common with that for a BEC . Indeed, integrating (\ref{Fermi1})
over $\textbf{R}$ produces, in $2d$, the momentum distribution $|\tilde{\phi}(\textbf{k}|^{2}=(a_{0}^{4}/4\pi)(k_{F}^{2}-k^{2})$,
where $k_{F}^{2}=2mE_{F}/\hbar^{2}$ and $a_{0}=(\hbar/m\omega)^{1/2}$
is the \textquotedbl{}oscillator size\textquotedbl{}. Similarly, integration
over $\textbf{k}$ results in an inverted parabola $(1/4\pi a_{0}^{4})(R_{F}^{2}-R^{2})$
for the initial density of the cloud $(R_{F}^{2}=2E_{F}/m\omega^{2})$.
Thus, both the density and the momentum distributions have the same
shape as for a BEC in a harmonic trap, see, respectively, (\ref{n1})
and (\ref{mom}). In $3d$ the shapes, although somewhat different,
are qualitatively similar. The treatment in \cite{Beilin} can be
easily extended to finite temperature $T$ if one uses in (\ref{AppF1}),
instead of the step function, the expression $\{exp[\beta(\frac{\hbar^{2}k^{2}}{2m}+\frac{1}{2}m\omega^{2}R^{2}-\mu)]+1\}^{-1}$,
where $\beta=1/k_{B}T$.

The problem of localization, as reviewed in Sec. 5.3, should apply,
with slight modifications, also to the Fermi gas. So far, however,
no specific calculations along these lines have been done.
On the experimental side, $3d$ localization of a Fermi gas has
been observed in \cite{Kondov}, where a gas of $^{40}K$ was
released from a harmonic trap into a speckle potential. According
to the authors of \cite{Kondov} their speckle potential is well
described by a anisotropic Gaussian autocorrelation function
(rather than by the more complicated function in Eq. (\ref{corr
gamma})). The gas was spin polarized, in order to make the
interatomic interactions negligible. The temperature range was
between 200 and 1500nK, well above the Fermi energy, so that
quantum degeneracy played no role. Upon release from the trap,
some part of the atomic cloud had diffused away and the remaining
localized component was observed. Again, as for the BEC
\cite{Jen}, the present state of the theory of Anderson
localization in a anisotropic $3d$ speckle potential does not
allow for a detail comparison with the experiment.

It is worthwhile to emphasize that the notion of the average particle
density $\overline{n}(\textbf{r},t)$ involves two kinds of averaging.
First, one performs the quantum (or thermal) average, i.e. writes
down the expectation value $\langle\hat{n}(\textbf{r},t)\rangle$
of the density operator $\hat{n}(\textbf{r},t)$, for a given realization
of the random potential. Then averaging over the statistical ensemble
of realizations is performed, yielding $\overline{\langle\hat{n}(\textbf{r},t)\rangle}\equiv\overline{n}(\textbf{r},t)$
\cite{footB} (In \cite{Beilin} this quantity is denoted as $n(\textbf{r},t)$,
without the overbar). One should keep in mind, as has been particularly
emphasized in \cite{Legg,Alt}, that in a single imaging experiment
(with sufficient resolution) one does not measure the expectation
value $\langle\hat{n}(\textbf{r},t)\rangle$ but rather one particular
event, i.e. some particular density pattern, $n(\textbf{r},t)$, whose
probability is dictated by the many-body wave function of the system.
Therefore even in the absence of disorder a single experimental image
will look \textquotedbl{}noisy\textquotedbl{} and grainy. (Only upon
averaging over many measurements, taken under identical experimental
conditions, will one obtain the quantum expectation value $\langle\hat{n}(\textbf{r},t)\rangle$.)
Fluctuations and correlations in such noisy density patterns are characterized
by the (equal time) correlation function $\langle\hat{n}(\textbf{r},t)\hat{n}(\boldsymbol{r}',t)\rangle$.
For a clean, homogeneous Fermi gas, in equilibrium, this correlation
function is described, for instance, in \cite{LLStat}. It exhibits
decaying oscillations, with a characteristic spatial period $\Delta x_{0}\sim k_{F}^{-1}$.
Such oscillatory behavior occurs also in a gas confined to a harmonic
trap. When the gas is released from the trap, it expands ballistically
(in the absence of disorder). The size of the atomic cloud grows linearly
with time and so does the correlation length $\Delta x(t)\approx\omega t\Delta x_{0}$.
(In $1d$, an exact expression for the correlation function of a freely
expanding Fermi gas has been obtained in \cite{ng}.) Thus, the free
(ballistic) expansion amplifies the scale of correlations.

It has been shown in \cite{Hens} that in the presence of a random
potential, i.e. when the expansion is diffusive instead of ballistic,
the picture is different: the size of the atomic cloud grows as $\sqrt{t}$
whereas the (short-range) correlations do not get amplified at all.
The authors studied the density-density correlation function \begin{eqnarray}
C(r,r',t) & = & \overline{\langle\hat{n}(r,t)\hat{n}(r',t)\rangle}\,-\,\overline{n}(r,t)
\overline{n}(r',t)\,-\,\delta(r-r')\overline{n}(r,t)\,,\quad\label{ddd}\end{eqnarray}
 where the last term describes trivial correlations, which exist already
in a classical ideal gas and which are commonly subtracted, in order
to isolate the non-trivial correlations \cite{LLStat}. Here we only
present an approximate analytic expression for the normalized correlation
function in a $3d$ gas (at zero temperature): \begin{eqnarray}
\frac{C(r,r',t)}{\overline{n}(\boldsymbol{r},t)\overline{n}(\boldsymbol{r}',t)}
 & = & -\frac{9}{g_{s}}e^{-\Delta r/l_{F}}\frac{\left[\sin(k_{F}\Delta r)\,-\,\left(k_{F}
 \Delta r\right)\cos(k_{F}\Delta r)\right]^{2}}{\left(k_{F}\Delta r\right)^{6}}\,,\label{3dcor}\end{eqnarray}
 where $g_{s}$ is the spin degeneracy factor, $l_{F}$ is the mean
free path at the Fermi energy, and $\Delta r=|\boldsymbol{r}-\boldsymbol{r}'|$.
The corresponding expression for the $2d$ case can be found in \cite{Hens}.

It is quite remarkable that, while the atomic cloud keeps expanding,
the normalized correlation function, (\ref{3dcor}), does not depend
on time. The characteristic length of oscillations remains the same
as for the gas in the trap, in sharp contrast with the ballistic case,
where the spatial oscillation period is growing linearly with time.
The peculiar behavior of correlations in a diffusively expanding gas
is related to phase randomization of the wave function of a diffusing
particle. Eq. (\ref{3dcor}) implies that, as the expansion proceeds,
the Fermi gas becomes less \textquotedbl{}rigid\textquotedbl{} \cite{Hens},
in the sense that the relative particle number fluctuation increases
and approaches the Poisson limit.

\section{Free expansion from a disordered trap}

So far we have discussed two types of problems: The disordered BEC
groundstate, with the ensuing insulator-superfluid transition (Sec.
4), and transport of cold atoms released from a trap into a random
potential (Sec. 5). In the present section we consider a different
set-up: The trapping potential $V_{tr}(\textbf{r})$ and the random
potential $V(\textbf{r})$ coexist in the same spatial region, and
the atoms are allowed to reach equilibrium in the combined potential.
Then, at $t=0$, \textit{both} potentials are switched off and the
atomic system undergoes a free expansion. We discuss in some detail
the case of a strongly anisotropic BEC, in a wave guide geometry,
and towards the end briefly mention some other cases. Our discussion
closely follows Ref. \cite{CBS}.

The BEC is initially strongly confined in the radial direction, ${\boldsymbol{\rho}}$,
by a harmonic trap of frequency $\omega_{\bot}$. Weak confinement
in the axial direction ($z$) is not essential for our considerations
and will be neglected. Thus, prior to its release the condensate is
in the ground state (we assume zero temperature), corresponding to
the radial confinement potential $\frac{1}{2}\omega_{\bot}^{2}\rho^{2}$
and a $z$-dependent potential $V(z)$. The latter can be random or
deterministic, with characteristic amplitude $V_{0}$ and scale of
variation $R_{0}\equiv1/k_{0}$. We assume a weak, smoothly varying
potential, in the sense that $V_{0}<<\mu$ and $k_{0}a_{\bot}<<1$,
where $a_{\perp}=\sqrt{2\mu/m\omega_{\perp}^{2}}$ is the radius of
the BEC in the trap. The chemical potential $\mu$ is assumed to be
much larger than $\hbar\omega_{\perp}$. Under these conditions the
Thomas-Fermi approximation is justified and the ground state density
is \begin{equation}
n_{0}(\rho,z)=\frac{1}{g}\left(\mu-V(z)-\frac{1}{2}m\omega_{\perp}^{2}\rho^{2}\right)\;.\label{C1}\end{equation}
 At time $t=0$ all potentials are switched off and the condensate
expands according to the equations (\ref{con}), (\ref{euler}). These
equations are to be solved with the initial conditions (\ref{C1})
for the density and $\textbf{v}=0$ for the velocity field. The condensate
rapidly expands in the radial direction but, due to the initial density
modulation, it also develops an axial velocity component $v_{z}(z,t)$
and the related phase. Upon completion of the first stage of the expansion,
at time of the order of $t_{0}=1/\omega_{\perp}$, the phase imprinted
on the condensate is \cite{LSP08} \begin{equation}
\theta(z)=\frac{\pi}{2\hbar\omega_{\perp}}V(z).\label{C2}\end{equation}

During the second stage of the expansion, for times $t>>t_{0}$, this
phase imprint can lead to large density mudulation, of the order or
larger than $\mu$, including the possible formation of matter wave
caustics. Since we are interested in large effects of this kind, we
can neglect from now on the weak initial density modulation (its crucial
part was to produce the phase imprint $\theta(z)$). The second stage
of the expansion amounts to linear evolution of the BEC wavefunction
with the imprinted phase. The wavefunction can be factorized into
radial and axial parts, $\Psi(\rho,z,t)=\Phi(\rho,t)\psi(z,t)$, and
we are interested in $|\psi(z,t)|^{2}$, which gives the density at
point $z$ and time $t$, normalized by the radial factor $|\Phi(\rho,t)|^{2}$.
The function $\psi(z,t)$ satisfies the linear, time-dependent Schrödinger
equation, with the initial condition $\psi(z,t_{0})={\rm e}^{{\rm i}\theta(z)}$,
whose solution is \begin{equation}
\psi(z,t)=\sqrt{\frac{m}{2\pi{\rm i}\hbar t}}\int{\rm d}z'\;\exp\left[\frac{{\rm i}m}{2\hbar t}(z-z')^{2}+{\rm i}\theta(z')\right]\;,\;\;\;\;\;\;\;\;(t>>t_{0})\label{C3}\end{equation}
 where the impressed phase $\theta(z)$ can be an arbitrary function
of $z$. The typical variation, $\theta_{0}=V_{0}/\hbar\omega_{\perp}$,
of the imprinted phase is the important parameter. In the case $\theta_{0}<<1$,
implicitly assumed in \cite{LSP08}, relative density variations remain
small for all times, as is clear from expanding the factor $\exp{\rm i}\theta(z')$
in the integrand of Eq~(\ref{C3}). Large effects, however, occur
in the opposite case, $1<<\theta_{0}<<\mu/\hbar\omega_{\perp}$ (the
latter inequality stems from the requirement $V_{0}<<\mu$).

We shall concentrate on the large $\theta_{0}$ case, and the
purpose is to find the form of the relative density
$|\psi(z,t)|^{2}$ at a given instant $t$. This gives the
$z$-dependence of the density of an expanding atomic cloud,
measured with a probe beam perpendicular to the condensate axis.
The problem is analogous to the phase screen model in optics,
studied extensively by Berry \cite{Berry}. In that
model a monochromatic plane wave encounters a thin screen. 
 The screen impresses on the wave a phase which may be deterministic
or random. For strong variation of this phase the wave passing through
the screen will develop large intensity variations. Observation of
the wave intensity at a point sufficiently far beyond the screen will
reveal a pattern of bright lines ( caustics). 
Although the arrangement we consider differs substantially from that
in optics (the corresponding matter wave is not at all monochromatic,
the mechanism of impression of the phase is different, and time assumes
the role of the spatial axis of propagation in optics), the mathematical
treatment of the two problems is essentially the same.

Before discussing the random case, it is instructive to consider a
sinusoidal modulation $\theta(z)=\theta_{0}\cos k_{0}z$ . We introduce
the dimensionless variables $\tilde{z}=k_{0}z$ and $\tilde{t}=t/t^{*}$,
with $t^{*}=m/\hbar k_{0}^{2}\theta_{0}$, and rewrite (\ref{C3})
as \begin{equation}
\psi(\tilde{z},\tilde{t})=\sqrt{\frac{\theta_{0}}{2\pi{\rm i}\tilde{t}}}\int{\rm d}\zeta\;\exp\left[{\rm i}\theta_{0}\varphi(\zeta,\tilde{z},\tilde{t})\right]\;,\label{C4}\end{equation}
 with \begin{equation}
\varphi(\zeta,\tilde{z},\tilde{t})=\frac{(\tilde{z}-\zeta)^{2}}{2\tilde{t}}+\cos\zeta\;.\end{equation}
 The relative density $|\psi(z,t)|^{2}$, initially unity, acquires
spatial variations with the passage of time. For $\tilde{t}\ll1$,
relative density modulations are small. They grow linearly with
$\tilde{t}$ and are oscillatory in $\tilde{z}$ with period $2\pi$.
Growth of density maxima with time culminates, for
$\theta_{0}\gg1$, in the formation of caustics at times
$\tilde{t}\gtrsim1$. Caustics are determined by the vanishing of
the first and second derivatives of the phase
$\varphi(\zeta,\tilde{z},\tilde{t})$ with respect to $\zeta$. The
first requirement defines rays of atoms, while the second
identifies the points at which the emerging rays are focused. The
density at these points is found by computing the integral in
Eq.~(\ref{C4}), using the stationary phase method. Since the first
two derivatives of $\varphi(\zeta,\tilde{z},\tilde{t})$ vanish,
the integral is controlled by the third derivative,
$\partial_{\zeta}^{3}\varphi(\zeta,\tilde{z},\tilde{t})$, and has
a value proportional to $\theta_{0}^{-1/3}$, resulting in a large
relative density,
$|\psi(\tilde{z},\tilde{t})|^{2}\sim\theta_{0}^{1/3}$, with the
characteristic for caustics aperiodic oscillations \cite{Berry}. A
numerical example is given in Fig. 5, where the relative density
is plotted as a function of $\tilde{z}$ for three different values
of $\tilde{t}$ (in this example $\theta_{0}=30$). Caustics are
located at $\tilde{z}=0$ for $\tilde{t}=1$ , and at
$\tilde{z}\approx\pm0.7$
for $\tilde{t}=2$.%
\begin{figure}[H]
\centering{}\includegraphics[scale=0.5]{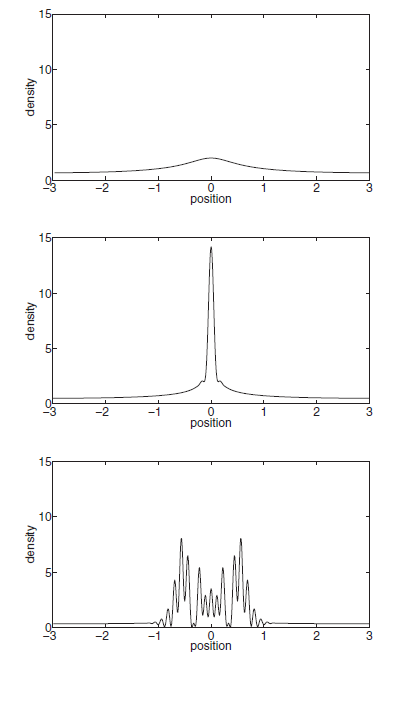}\caption{Relative
density as a function of (dimensionless) position $\tilde{z}$, for
$\theta_0=30$ and   $\tilde{t}=0.5$ (top), 1 (middle) and 2
(bottom). (Reprinted, with permission, from Phys. Rev. A
\textbf{80} 013603 (2009)).}
\end{figure}

Eq.~(\ref{C3}) for $\psi(z,t)$ applies also for a disordered potential
and the mechanism for caustic formation is qualitatively the same
as for the sinusoidal potential discussed above. Since the impressed
phase $\theta(z')$ is now a random function, the formation of caustics
is a matter of probability. The typical time for caustic formation
is $t^{*}=mR_{0}^{2}/\hbar\theta_{0}$. Using the relations $\frac{1}{2}m\omega_{\perp}^{2}a_{\perp}^{2}=\mu$
and $\theta_{0}=\pi V_{0}/2\hbar\omega_{\perp}$, the formation time
can be written in terms of the experimentally controlled parameters
as \begin{equation}
t^{*}=\frac{4}{\pi\omega_{\perp}}\frac{\mu}{V_{0}}\left(\frac{R_{0}}{a_{\perp}}\right)^{2}\;.\end{equation}
 Quantitative analytic results for the second moment of the relative
density, $\overline{|\psi(z,t)|^{4}}\equiv S(t)+1$, can be found
in \cite{CBS}. The calculation is along the same lines as for the
random screen problem in optics \cite{Jake}, where $S$ is called
the scintillation index and it is a measure of spatial intensity fluctuations
in the speckle pattern. For a uniform intensity $S$ is clearly zero,
and for the standard, fully developed speckle pattern $S$ is unity.
These two limits correspond, respectively, to short ( $\tilde{t}<<1$)
and long ( $\tilde{t}>>1$) times. In the more interesting intermediate
time regime, when $\tilde{t}\gtrsim1$, $S$ is proportional to $\ln\theta_{0}>>1$,
signalling the appearance at this time of caustics and the associated
large density fluctuations.

Large density variations in an expanding condensate, released from
a disordered trap, were observed in \cite{Chen}. In this experiment
$\mu/\hbar\omega_{\perp}=5.6$, $R_{0}=15\mu{\rm m}$, $a_{\perp}\sim10\mu{\rm m}$
and the largest density variations were observed for $V_{0}=0.5\mu$,
in a time of flight image at $t_{{\rm ToF}}=8{\rm ms}$, which is
significantly larger than $1/\omega_{\perp}$, see Fig. 6(h). %
\begin{figure}[H]
\centering{}\includegraphics[scale=0.5]{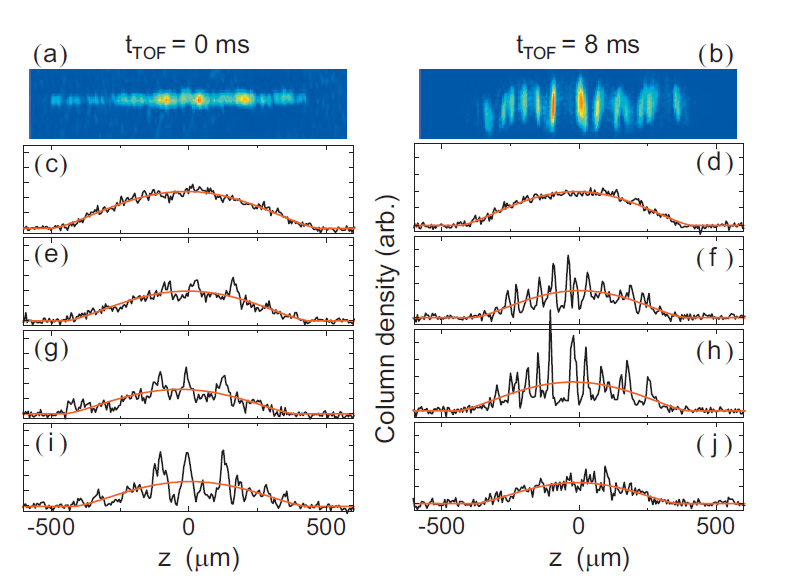}\caption{(c)-(j) {\it
In situ} column density profiles on left and corresponding ToF
profiles on right: (c),(d) $V_0=0$; (e),(f) $V_0=0.3\mu$; (g),(h)
$V_0=0.5\mu$; and (i),(j) $V_0=1.0\mu$. Solid red lines are fits
to Thomas-Fermi distributions. (Reprinted, with permission, from
Phys. Rev. A \textbf{77}, 033632 (2008)).}

\end{figure}
The value $V_{0}=0.5\mu$ corresponds to a phase amplitude $\theta_{0}=4.4$
which gives the caustic formation time $t^{*}=5{\rm ms}$, comparable
to the value of $t_{{\rm ToF}}$. While the value $\theta_{0}=4.4$
does not lie deep within the large $\theta_{0}$ regime ( the experiment
was not designed for caustic observation), the large density variations
in Fig.~6(h) can most likely be attributed to caustics. On the other
hand, a similar experiment performed in \cite{LSP08} cannot be interpreted
in terms of caustics, as discussed in \cite{CBS}.

So far our discussion was limited to a strongly anisotropic, quasi-$1d$
BEC. Let us briefly comment on some other cases, referring the reader
to \cite{CBS} for details.

\textit{(i) $2d$ geometry}. A quasi-$2d$ condensate is initially
strongly confined in the axial $z$-direction (trap frequency
$\omega_{z}$), with only a weak confinement in the radial ${\bf
r}_{\perp}$-direction. In addition, there is a random or a
deterministic in-plane potential $V({\bf r}_{\perp})$. After all
potentials are switched  off and the condensate is freely
expanding for a time $\sim1/\omega_{z}$, it acquires a phase
imprint $\theta({\bf r}_{\perp})\sim V({\bf
r}_{\perp})/\hbar\omega_{z}$. At later times, in analogy with
Eq.~(\ref{C3}), the planar part of the condensate wave function is
given by \begin{equation} \Phi({\bf
r}_{\perp},t)=\frac{m}{2\pi{\rm i}\hbar t}\int{\rm d}^{2}{\bf
r}_{\perp}^{\prime}\exp\left[\frac{{\rm i}m}{2\hbar t}|{\bf
r}_{\perp}-{\bf r}_{\perp}^{\prime}|^{2}+{\rm i}\theta({\bf
r}_{\perp}^{\prime})\right]\;.\label{planar-phi}\end{equation}
 As in quasi-one dimensional systems, caustics are formed for $\theta_{0}\gg1$
at values of the scaled time $\tilde{t}\sim1$, and in this regime
Eq.~(\ref{planar-phi}) can be evaluated using the stationary phase
method. Unlike the quasi-$1d$ case, now caustics form a set of lines
rather than a set of isolated points. The theory of caustic formation
resulting from a random phase $\theta({\bf r}_{\perp})$ is analogous
to the treatment of the phase screen problem, which has been studied
extensively for two-dimensional systems in the context of optics \cite{Berry}.
In particular, the morphology of caustic lines for the random case
is discussed in \cite{Berry1}.

\textit{(ii) interacting bosons in $1d$ }. Here the transverse confinement
is very tight: $\hbar\omega_{\perp}$ is much larger than the characteristic
interaction energy so that, with respect to transverse motion, all
$N$ atoms in the trap reside in the ground state $\chi_{0}(\rho)$
of the harmonic oscillator, forming a strictly one-dimensional system.
The axial motion is controlled by an effective one-dimensional Hamiltonian
-- the Lieb-Liniger model \cite{lieb,olshanii}. At some instance
the radial confinement is switched off and, simultaneously, a short
potential pulse is applied to the system, creating a prescribed phase
imprint $\theta(z)$ which can be a deterministic or a random function
of $z$. It has a characteristic amplitude $\theta_{0}$ and scale
of variation $R_{0}$. The system is beyond mean field theory and
the whole many-body wave function has to be considered. Initially,
i.e. just after switching off the trap and making the phase imprint,
the axial part of the many-body function is \begin{equation}
\Psi(z_{1},\ldots z_{N};t{=}0)=\exp\left[{\rm i}\sum_{j=1}^{N}\theta(z_{j})\right]\Phi_{0}(z_{1},\ldots z_{N})\label{initial-1d}\end{equation}
 where $\Phi_{0}(z_{1},\ldots z_{N})$ is the ground state wavefunction,
prior to the action of the pulse. In the course of expansion, large
density variations can develop, including formation of caustics. One
must, however, account for the momentum distribution function $n(p)$
of the system of interacting bosons (in the mean field approach $n(p)=2\pi\delta(p)$).
It is clear that a broad momentum distribution will impede focusing
of the atomic rays and, thus, formation of caustics. The width of
the distribution $n(p)$ depends on the strength of interactions and,
roughly, it is determined by the inverse coherence length $1/\xi_{c}$.
It is shown in \cite{CBS} that the necessary condition for caustics
is $\xi_{c}n_{1}\gg1$, where $n_{1}$ is the one-dimensional density.
This condition can be satisfied only in the weak interaction limit.
In the opposite limit of strong interactions (hard core bosons) $\xi_{c}\approx1/n_{1}$
and the caustics are suppressed.

\textit{(iii) $1d$ Fermi gas}. Similarly to bosons, one can consider
the dynamics of fermions, with an initially impressed phase that varies
periodically or randomly as a function of position $z$. For a $1d$
Fermi gas the problem is similar to that of the previous item. The
particle statistics manifests itself only through the momentum distribution
of particles before the expansion, and the Fermi wavelength $\lambda_{{\rm F}}$
is taking the place of the coherence length $\xi_{c}$ of the Lieb-Liniger
gas. Since $\lambda_{{\rm F}}n_{1}\sim1$, caustics are absent from
the Fermi gas, for the same reason as in the Bose gas with hard-core
interactions. The main difference between non-interacting fermions
and hard-core bosons stems from the Fermi surface discontinuity in
the momentum distribution. Within the approximations of geometrical
optics, this discontinuity leads to sharp peaks in the derivative
of the density (with respect to position $z$ or time $t$), rather
than in the density itself.

\section{Conclusions}

Several topics concerning the behavior of cold atomic gases in the
presence of disorder have been reviewed. This is a relatively new
and rapidly progressing subject which poses some questions, not previously
encountered in condensed matter physics. One such question pertains
to Anderson localization in a $3d$ speckle potential, with its rather
unusual anisotropic, long-range correlations. It is not yet clear
how these peculiarities affect the standard picture of Anderson localization
and work in this direction has just begun \cite{Kondov,Jen,Pira}.
Even such basic quantity as the average density of states is not well
known for this kind of a random potential and the problem awaits careful
theoretical investigation.

In the experimental setup of \cite{Kondov,Jen} one observes the
time evolution of an atomic cloud (BEC or fermions) in the
presence of a random potential. In the long time limit, part of
the cloud diffuses away and the rest remains localized. To obtain
an accurate value for the localization length and for the exponent
$\nu$ one has to measure the long tail of the localized part (or
the very slowly diffusing part, originating form the energy
components close to $E_{c}$). So far such measurements are beyond
reach. Another possibility, mentioned in \cite{Jen}, would be to
prepare a collection of atoms with a narrow energy distribution
and to study its evolution in a random potential. This would allow
to \textquotedbl{}isolate\textquotedbl{} the transition region
from the rest of the spectrum.

One of the most challenging problems in condensed matter physics
is the combined effect of interactions and disorder. Since
non-interacting bosons correspond to to a singular, pathological
limit, it is clear that the interactions must be accounted for
from the start. Already the Gross-Pitaevskii mean field theory for
disordered bosons leads to the interesting and difficult problem
of the $2d$ and $3d$ non-linear Schrödinger equation with a random
potential. Two of the recently addressed theoretical problems in
this context are the time evolution of a \textquotedbl{}matter
wave packet\textquotedbl{} \cite{Cher,Schwiete,Cher1} and the flow
of a BEC through a disordered region \cite{Paul}. Things get more
involved when one tries to go beyond mean field. Even the nature
of the ground state, and the zero-temperature superfluid-insulator
transition, is only qualitatively understood \cite{Shk,Fal}.
Finite temperature introduces a further dimension to the problem
and allows for a number of phase transitions. In particular, it
has been argued \cite{Al1} that in $2d$ no direct
superfluid-insulator transition is possible at finite temperature
and that, under increase of disorder, the system of weakly
interacting bosons must pass through a normal fluid state. (In
$3d$ the sharp fluid-insulator transition is replaced by a
crossover). The combined effect of interactions, disorder and
temperature ( beyond the scope of this review) will surely receive
much attention in the future.

\section{Acknowledgments}
I am indebted to  L. Beilin, J. Chalker, E. Gurevich, P. Henseler,
A. Minguzzi, S. Skipetrov and B. van Tiggelen  for collaboration
on some of the topics presented in this review. I am grateful to
J. Chalker, E. Gurevich, S. Lipson, C. M\"{u}ller, D. Polyakov, M.
Raikh, M. Segev and S. Skipetrov for extensive discussions and
correspondence which led to clarification of a number of specific
topics addressed in this review. Useful conversations and
correspondence with I. Aleiner, B. Altshuler,  A. Aspect, N.
Cherroret,  J. Feinberg, A. Genack, F. Jendrzejewski, S. Kondov,
B. DeMarco, C. Miniatura, M. Piraud and L. Sanchez-Palencia are
gratefully acknowledged.



\begin{thebibliography}{97}
\bibitem{AI} Aspect A and Inguscio M 2009 Phys. Today \textbf{62}
30

\bibitem{SP1} Sanchez-Palencia L and Lewenstein M 2010 Nature Phys.
\textbf{6}, 87

\bibitem{Fall} Fallani L, Fort C and Inguscio M 2008 \textit{Advances
in Atomic, Molecular and Optical Physics} \textbf{56} 119

\bibitem{Mod1} Modugno G 2010 Rep. Prog. Phys. \textbf{73} 102401

\bibitem{Lew} Lewenstein M, Sanpera A, Ahufinger V, Damski B, Sen
(De) A, and Sen U 2007 Adv. Phys. \textbf{56} 243

\bibitem{Cord} Müller C A and Delande D 2009 \textit{Ultracold Gases
and Quantum Information}, Les Houches XCI, p.441

\bibitem{Billy} Billy J, Josse V, Zuo Z, Bernard A, Hambrecht B,
Lugan P, Clément D, Sanchez-Palencia L, Bouyer P and Aspect A 2008
\textit{Nature} \textbf{453} 891

\bibitem{Roati} Roati G, D'Errico C, Fallani L, Fattori M, Fort C,
Zaccanti M, Modugno G, Modugno M and Inguscio M 2008 \textit{ Nature}
\textbf{453} 895

\bibitem{MR} Robert-de-Saint-Vincent M, Brantut J.-P., Allard B,
Plisson T, Pezzé L, Sanchez-Palencia L, Aspect A, Bourdel T, and Bouyer
P 2010 Phys. Rev. Lett. \textbf{104} 220602

\bibitem{Kondov} Kondov S S, McGehee W R, Zirbel J J, and DeMarco
B 2011 Science \textbf{334} 66

\bibitem{Jen} Jendrzejewski F, Bernard A, Müller K, Cheinet P, Josse
V, Piraud M, Pezzé L, Sanchez-Palencia L, Aspect A, and Bouyer P 2011
arXiv:1108.0137

\bibitem{String} Pitaevskii L and Stringari S 2003 \textit{Bose-Einstein
Condensation}, Clarendon Press

\bibitem{Pet} Pethick C J and Smith H 2002 \textit{Bose-Einstein
Condensation in Dilute Gases}, Cambridge University Press

\bibitem{Clement1} Clément D, Varon A F, Hugbart M, Retter J A, Bouyer
P, Sanchez-Palencia L, Gangardt D M, Shlyapnikov G V, and Aspect A
2005 Phys. Rev. Lett. \textbf{95} 170409

\bibitem{Clement2} Clément D, Varon A F, Retter J A, Sanchez-Palencia
L, Aspect A, and Bouyer P 2006 New. J. Phys. \textbf{8} 165

\bibitem{Fort} Fort C, Fallani L, Guarrera V, Lye J E, Modugno M,
Wiersma D S, and Inguscio M 2005 Phys. Rev. Lett. \textbf{95}, 170410

\bibitem{Chen} Chen Y P, Hitchcock J, Dries D, Junker M, Welford
C, and Hulet R G 2008 Phys. Rev. A \textbf{77} 033632




\bibitem{Goodman} Goodman J W 2007 Speckle Phenomena in Optics, Roberts
and Company, Englewood, Colorado

\bibitem{giglio} Giglio M, Carpineti M and Vailati A 2000 Phys. Rev.
Lett. \textbf{85} 1416

\bibitem{cerbino} Cerbino R 2007 Phys. Rev. A \textbf{75} 053815

\bibitem{gatti} Gatti A, Magatti D and Ferri F 2008 Phys. Rev. A
\textbf{78} 063806

\bibitem{magatti} Magatti D, Gatti A and Ferri F 2009 Phys. Rev.
A \textbf{79} 053831

\bibitem{lipson} Lipson A, Lipson S G and Lipson H S 2010 \textit{Optical
Physics}, Cambridge

\bibitem{goodman1} Goodman J W 2005 Introduction to Fourier Optics,
Roberts and Company, Englewwod, Colorado 
\bibitem{Note1}see Eq. (17) in \cite{magatti}, with
$z_{1}\equiv z$ and with the ratio $z_{1}/z_{2}$ in the exponent
set equal to unity.

\bibitem{Kuhn} Kuhn R C, Sigwarth O, Miniatura C, Delande D and Müller
C A 2007 New Jour. Phys. \textbf{9} 161

\bibitem{Mini} Miniatura C, Kuhn R C, Delande D, and Müller C A 2009
Eur. Phys. J. B \textbf{68}  353

\bibitem{Pezze} Pezzé L, Robert-de-Saint-Vincent M, Bourdel T, Brantut
J.-P., Allard B, Plisson T, Aspect A, Bouyer P, and Sanchez-Palencia
L 2011 New Jour. Phys. \textbf{13} 095015

\bibitem{Gavish} Gavish U and Castin Y 2005 Phys. Rev. Lett. \textbf{95}
020401

\bibitem{Gad} Gadway B, Pertot D, Reeves J, Vogt M, Schneble D 2011
Phys. Rev. Lett. \textbf{107} 145306

\bibitem{ES} Shklovskii B I and Efros A L 1984 \textit{Electronic
Properties of Doped Semiconductors }, Springer

\bibitem{L} Lifshits I M, Gredeskul S A and Pastur L A 1988 \textit{Introduction
to the Theory of Disordered Systems}, Wiley

\bibitem{Lee} Lee P A and Ramakrishnan T V 1985 Rev. Mod. Phys. \textbf{57}
287

\bibitem{Kram} Kramer B and MacKinnon A 1993 Rep. Prog. Phys. \textbf{56}
1469

\bibitem{Akker} E. Akkermans and G. Montambaux 2007 \textit{Mesoscopic
Physics of Electrons and Photons}, Cambridge University Press

\bibitem{Efetov} Efetov K 1997 \textit{Supersymmetry in Disorder
and Chaos}, Cambridge

\bibitem{Mirlin} Mirlin A D 2000 Phys. Rep \textbf{326} 259

\bibitem{Voll} Vollhardt D and Wölfle p 1992 in \textit{Electronic
phase transitions} Eds. W. Hanke and Y. V. Kopaev (Elsevier, Amsterdam)

\bibitem{Shk} Shklovskii B I 2008 Semiconductors (St. Petersburg)
\textbf{42} 927

\bibitem{Fal} Falco G M, Natterman T and Pokrovsky V L, 2009 Phys.
Rev. B \textbf{80} 104515

\bibitem{And} Anderson P W 1958 Phys. Rev. \textbf{109} 1492 


\bibitem{LL} Landau L D and Lifshitz E M 1997 Quantum Mechanics (Third
Edition), Butterworth - Heinemann.

\bibitem{Slevin}Slevin K and Ohtsuki T 1999 Phys. Rev. Lett. \textbf{82}
382 


\bibitem{Skip} Skipetrov S E, Minguzzi A, van Tiggelen B A and Shapiro
B 2008 Phys. Rev. Lett. \textbf{100}  165301

\bibitem{Tig} Yedjour A and Van Tiggelen B A 2010 Eur. Phys. J. D
\textbf{59} 249




\bibitem{H} Halperin B I and Lax M 1966 Phys. Rev. \textbf{148} 722

\bibitem{Z} Zittartz J and Langer J S 1966 Phys. Rev. \textbf{148}
741

\bibitem{fal1} Falco G M, Fedorenko A A, Giacomelli J and Modugno
M 2010 Phys. Rev. A \textbf{82} 053405

\bibitem{Pira} Piraud M, Pezzé L and Sanchez-Palencia L 2011 arXiv:1112.2859

\bibitem{DK}M. I. Dyakonov and A. V. Khaetskii 1991 Sov. Phys. JETF
\textbf{72}  590


\bibitem{foot1} The diffusion coefficient here should be inderstood
as averaged over various initial positions of the diffusing particle.

\bibitem{Zal}Zallen R 1983 \textit{The Physics of Amorphous Solids},
Wiley \& Sons

\bibitem{Per}Shapiro B 1983 \textquotedbl{}Localization versus percolation\textquotedbl{},
in \textit{Percolation Structures and Processes}, p. 367, Eds. G.
Deutscher, R. Zallen and J. Adler (Bristol: Adam Hilger)

\bibitem{Pil} Pilati S, Giorgini S, Modugno M, and Prokof'ev N 2010
New J. Phys. \textbf{12} 073003

\bibitem{Ab} Abrahams E, Anderson P W, Licciardello D C and Ramakrishnan
T V 1979 Phys. Rev. Lett. \textbf{42} 673

\bibitem{Raikh} Apalkov V M, Raikh M E and Shapiro B 2003 in \textit{\textquotedbl{}Anderson
Localization and its Ramifications\textquotedbl{}} Springer, 'Lecture
Notes in Physics', ed. T. Brandes and S. Kettermann, p. 119

\bibitem{BS1} Shapiro B 2007 Phys. Rev. Lett. \textbf{99}, 060602

\bibitem{Pollak} Pollak M and Ortuno M 1985 \textit{Electron-electron
interactions in disordered systems} eds. Efros A L and Pollak M, North-Holland,
287

\bibitem{Rep}Reppy J D 1992 J. Low Temp. Phys. \textbf{87} 205

\bibitem{Pas}Pasienski M, McKay D, White M and DeMarco B 2010 Nature
Physics \textbf{6} 667

\bibitem{Dei}Deissler B, Zaccanti M, Roati G, D'Errico C, Fattori
M, Modugno M, Modugno G, and Ignuscio M 2010 Nature Physics \textbf{6}
354

\bibitem{Gau} Gaul C and Müller C A 2011 Phys. Rev. A \textbf{83}
063629

\bibitem{Lug} Lugan P and Sanchez-Palencia L 2011 Phys. Rev. A \textbf{84}
013612

\bibitem{Al1} Aleiner I L, Altshuler B L and Shlyapnikov G V 2010
Nature Physics \textbf{6} 900

\bibitem{Huang} Huang K and Meng H-F 1992 Phys. Rev. Lett. \textbf{69}644

\bibitem{Gior} Giorgini S, Pitaevskii L and Stringari S 1994 Phys.
Rev. B \textbf{49}12938




\bibitem{DKKL} Lee D K K and Gunn J M F 1990 J. Phys. Cond. Matter
\textbf{2} 7753

\bibitem{Kag}Kagan Yu, Surkov E L and Shlyapnikov G V 1996 Phys.
Rev. A\textbf{54} R1753

\bibitem{Cas}Castin Y and Dum R 1996 Phys. Rev. Lett. \textbf{77}
5315

\bibitem{SP2} Sanchez-Palencia L, Clément D, Lugan P, Bouyer P, Shlyapnikov
G V, and Aspect A 2007 Phys. Rev. Lett. \textbf{98} 210401

\bibitem{Chalk} Chalker J T 1990 Physica A \textbf{167} 253




\bibitem{Hens} Henseler P and Shapiro B 2008 Phys.Rev. A \textbf{ 77}
033624

\bibitem{Cher2} Cherroret N and Skipetrov S E 2008 Phys. Rev. Lett.
\textbf{101} 190406

\bibitem{Cher} Cherroret N and Skipetrov S E 2009 Phys. Rev. A \textbf{79}
063604

\bibitem{Schwiete} Schwiete G and Finkel'stein A M 2010 Phys. Rev.
Lett. \textbf{104}  103904

\bibitem{Cher1} Cherroret N and Wellens T 2011 Phys. Rev. A \textbf{84}
021114

\bibitem{Weg} Wegner F 1976 Z. Phys. B\textbf{25} 327 


\bibitem{Fish} Fishman S, Grempel D R and Prange R E 1982 Phys. Rev.
Lett. \textbf{49} 509

\bibitem{Raiz} Moore F L, Robinson J C, Bharucha C F, Sundaram B and Raizen M G 1995 Phys. Rev.
Lett. \textbf{75} 4598

\bibitem{Cass} Casati G, Guarneri I and Shepelyansky D L 1989 Phys.
Rev. Lett. \textbf{62} 345

\bibitem{Del1} Chabé J, Lemarié G, Grémaud B, Delande D, Szriftgiser
P and Garreau J-C 2008 Phys. Rev. Lett. \textbf{101} 255702

\bibitem{Del2} Lemarié G, Chabé J, Szriftgiser P, Garreau J-C, Grémaud
B and Delande D 2009 Phys. Rev. A \textbf{80} 043626

\bibitem{Del3} Lopez M, Clément J-F, Szriftgiser P, Garreau J-C and
Delande D 2011 arXiv:1108.0630

\bibitem{Giorgini}  Giorgini S,  Pitaevskii L P and  Stringari S
2008 Rev. Mod. Phys. \textbf{80} 1215

\bibitem{YCastin} Y. Castin, in {}``Ultra-cold Fermi gases'', Proc.
Inter. School of Physics {}``Enrico Fermi'', Varenna, Eds. M. Inguscio,
W. Ketterle and C. Salomon, p.289 (2007).

\bibitem{Beilin} Beilin L, Gurevich E and Shapiro B 2010 Phys. Rev.
A \textbf{81} 033612

\bibitem{footB} In the mean field Gross-Pitaevskii theory, with its
\textquotedbl{}classical\textquotedbl{} macroscopic wave function,
the quantum average is implicit and only averaging over the realization
of disorder remains.

\bibitem{Legg} Leggett A J 2001 Rev. Mod. Phys. \textbf{73} 307

\bibitem{Alt} Altman E, Demler E and Lukin M D 2004 Phys. Rev. A
\textbf{70}, 013603

\bibitem{LLStat} Landau L D and Lifshitz E M 1980 Statistical Physics,
part 1, Butterworth- Heinemann

\bibitem{ng} Nagornykh P and Galitski V 2007 Phys. Rev. A \textbf{75}
065601

\bibitem{CBS} Chalker J T and Shapiro B 2009 Phys. Rev. A \textbf{80}
013603

\bibitem{LSP08} Clément D, Bouyer P, Aspect A, and Sanchez-Palencia
L 2008 Phys. Rev. A \textbf{77} 033631

\bibitem{Berry} Berry M V 1977 J. Phys. A \textbf{10} 2061


-------------1976 Adv. Phys. \textbf{25} 1

\bibitem{Jake} Jakeman E and McWhirter J G 1977 J. Phys. A
\textbf{10} 1599

\bibitem{Berry1} Berry M V, 1980 Proc. Symp. App. Maths \textbf{36}
13.

\bibitem{lieb} Lieb E H and Liniger W 1963 Phys. Rev. \textbf{130}
1605

\bibitem{olshanii} Olshanii M 1998 Phys. Rev. Lett. \textbf{81} 938

\bibitem{Paul} Paul T, Albert M, Schlagheck P, Leboeuf P and Pavloff
N 2009 Phys. Rev. A \textbf{80} 033615











































\end{thebibliography}
\end{document}